\documentclass[fleqn,usenatbib]{mnras}

\usepackage{newtxtext,newtxmath}
\usepackage[T1]{fontenc}
\usepackage{ae,aecompl}

\usepackage{graphicx}	% Including figure files
\usepackage{amsmath}	% Advanced maths commands
\usepackage{amssymb}	% Extra maths symbols
\usepackage{comment}
\usepackage[dvipsnames]{xcolor}
\usepackage{url}
\usepackage{subcaption}
\usepackage{natbib}

\newcommand{\hi}{H\,\textsc{i}}
\newcommand{\feiilong}{[Fe\,\textsc{ii}]$_{1.644\,\rm \mu m}$}
\newcommand{\feii}{[Fe\,\textsc{ii}]}
\newcommand{\hh}{H\textsubscript{2}}
\newcommand{\hco}{HCO\textsuperscript{+}}
 
\defcitealias{SutherlandBicknell2007}{SB07}
\defcitealias{GarciaBurillo2007}{Ga07}
\defcitealias{Perlman2001}{Pe01}
\defcitealias{Giroletti2003}{Gi03}

\title[Jet-ISM interactions in 4C\,31.04]{Jets blowing bubbles in the young radio galaxy 4C\,31.04}

\author[H. R. M. Zovaro et al.]{
Henry R. M. Zovaro$^{1}$\thanks{E-mail: henry.zovaro@anu.edu.au},
Robert Sharp$^{1}$,
Nicole P. H. Nesvadba$^{2}$,
Geoffrey V. Bicknell$^{1}$,
\newauthor
Dipanjan Mukherjee$^{3}$,
Alexander Y. Wagner$^{4}$,
Brent Groves$^{1}$,
Shreyam Krishna$^{1}$
\\
$^{1}$Research School of Astronomy and Astrophysics, The Australian National University, Canberra, ACT 2611, Australia\\
$^{2}$Institut d'Astrophysique Spatiale, UMR 8617, Universit\'e Paris-Sud, B\^{a}t. 121, 91405 Orsay, France \\
$^{3}$Dipartimento di Fisica Generale, Universita degli Studi di Torino, Via Pietro Giuria 1, 10125 Torino, Italy\\
$^{4}$University of Tsukuba, Center for Computational Sciences, Tennodai 1-1-1, 305-0006, Tsukuba, Ibaraki, Japan\\
}

\date{Accepted 2019 January 17. Received 2019 January 17; in original form 2018 November 22}

\pubyear{2019}

\begin{document}
\label{firstpage}
\pagerange{\pageref{firstpage}--\pageref{lastpage}}
\maketitle

% This is a simple template for authors to write new MNRAS papers.
% The abstract should briefly describe the aims, methods, and main results of the paper.
% It should be a single paragraph not more than 250 words (200 words for Letters).
% No references should appear in the abstract.
\begin{abstract}
We report the discovery of shocked molecular and ionised gas resulting from jet-driven feedback in the low redshift ($z = 0.0602$) compact radio galaxy 4C\,31.04 using near-IR imaging spectroscopy.
4C\,31.04 is a $\sim 100$\,pc double-lobed Compact Steep Spectrum source believed to be a very young AGN. It is hosted by a giant elliptical with a $\sim 10^{9}\,\rm M_\odot$ multi-phase gaseous circumnuclear disc.
We used high spatial resolution, adaptive optics-assisted $H$- and $K$-band integral field Gemini/NIFS observations to probe (1) the warm ($\sim 10^3$\,K) molecular gas phase, traced by ro-vibrational transitions of \hh{}, and (2), the warm ionized medium, traced by the \feiilong{} line.
The \feii{} emission traces shocked gas ejected from the disc plane by a jet-blown bubble $300-400\,\rm pc$ in diameter, whilst the \hh{} emission traces shock-excited molecular gas in the interior $\sim 1\,\rm kpc$ of the circumnuclear disc.
Hydrodynamical modelling shows that the apparent discrepancy between the extent of the shocked gas and the radio emission can occur when the brightest regions of the synchrotron-emitting plasma are temporarily halted by dense clumps, whilst less bright plasma can percolate through the porous ISM and form an energy-driven bubble that expands freely out of the disc plane.
Simulations suggest that this bubble is filled with low surface-brightness plasma not visible in existing VLBI observations of 4C\,31.04 due to insufficient sensitivity.
Additional radial flows of jet plasma may percolate to $\sim$ kpc radii in the circumnuclear disc, driving shocks and accelerating clouds of gas, giving rise to the \hh{} emission. 
\end{abstract}

\begin{keywords}
galaxies: individual: 4C\,31.04 -- galaxies: active -- galaxies: jets -- galaxies: nuclei -- ISM: jets and outflows -- ISM: kinematics and dynamics 
\end{keywords}

%%%%%%%%%%%%%%%%%%%%%%%%%%%%%%%%%%%%%%%%%%%%%%%%%%

%%%%%%%%%%%%%%%%% BODY OF PAPER %%%%%%%%%%%%%%%%%%

\section{Introduction}
 
% AGN feedback
% Only need 1-2 sentences
Feedback processes involving active galactic nuclei (AGN) have long been known to be important drivers of galaxy evolution. 
Quasar winds and jets from powerful AGN are believed to be important in shaping the galaxy luminosity function and in establishing the observed correlations between the properties of the bulge and the supermassive black hole\,\citep[][and references therein]{Silk&Rees1998,Tremaine2002,Croton2006,King&Pounds2015}. 
On much smaller scales, AGN feedback processes are likely to be equally important: in particular, interactions between radio jets and the interstellar medium (ISM) on sub-kpc scales may have a significant impact upon the evolution of the host galaxy, particularly in the earliest stages of jet evolution.

Hydrodynamical simulations of jets propagating through an inhomogeneous ISM\,\citep{Mukherjee2016,Wagner2016} have shown that star formation in the host galaxy can both be enhanced and inhibited by interactions between the jets and the ISM on sub-kpc scales.
\citet[][henceforth \citetalias{SutherlandBicknell2007}]{SutherlandBicknell2007} demonstrated that the evolution of young radio galaxies can be separated into distinct stages: a `flood-and-channel' phase, followed by the formation of an energy-driven bubble that creates a bow shock as it expands, after which the jet breaks free of the bubble, finally forming extended FR\,II-like lobes. 
The expanding bubble driven by the jet plasma can ablate clouds and accelerate them to high velocities, preventing star formation and driving powerful outflows\,\citep[e.g., in 3C\,326\,N;][]{Nesvadba2010}.
\citet{Mukherjee2016} found that the energy-driven bubble can remain confined to the galaxy's potential for a long time due to interactions with the inhomogeneous ISM. The bubble drives shocks and turbulence into the ISM, potentially leading to quenching of star formation.
Conversely, the over-pressured plasma in the hot bubble can trigger gravitational instabilities and perhaps cloud collapse, enhancing star formation\,\citep{Gaibler2012,Fragile2017}. 
Despite mounting evidence from simulations of such `positive feedback', jet-induced star formation has only been observed in a handful of sources, e.g., the $z = 3.8$ radio galaxy 4C\,41.17\,\citep{Bicknell2000}, 3C\,285\,\citep{Salome2015}, Centaurus A\,\citep{Salome2017} and in Minkowski's Object\,\citep{Salome2015,Lacy2017}.
Simulations show these feedback mechanisms are sensitive to both the ISM structure and jet power, making it difficult to predict whether star formation will be enhanced or inhibited, and in turn the impact on the evolution of host galaxy\,\citep{Wagner2016,Zubovas&Bourne2017,Mukherjee2018b}. 
High-resolution observations of the ISM in young radio galaxies are therefore key to exposing the relationship between the properties of the radio jets and the host galaxy. 

% GPS and CSS sources.
Gigahertz Peak Spectrum (GPS) and Compact Steep Spectrum (CSS) sources are extragalactic radio sources characterised by a peak in their radio spectrum occurring in the GHz and 100\,MHz range for GPS and CSS sources respectively and compact ($< 1$\,kpc) radio emission with resolved lobes and/or jets\,\citep[for a comprehensive review see][]{O'Dea1998}. 
Recent observations \citep[e.g.,][]{Tingay2015,Callingham2017} indicate that the spectral peak is most likely caused by free-free absorption (FFA) of synchrotron emission by an ionized ISM with a varying optical depth, as proposed by \citet{Bicknell1997}. 
More recently, hydrodynamical simulations by \citet{Bicknell2018} have indeed demonstrated that jets percolating through an ionized, inhomogeneous ISM can reproduce the observed spectra of GPS/CSS sources via FFA. 
The peaked spectrum and compact size of GPS and CSS sources suggests that they harbour young jets, temporarily confined by a dense ISM, and that the more powerful GPS ans CSS sources are the progenitors of classical double-lobed radio sources.
This `youth hypothesis' is supported by age estimates based on breaks in the synchrotron spectrum and on hotspot advance velocities\,\citep{O'Dea1998}.
The compact nature of the jets in GPS and CSS sources therefore enables us to study jet-ISM interactions within the host galaxy in the earliest stages of evolution, and therefore represent an important class of sources in the context of AGN feedback. 

% The goal of this paper and why we study 4C\,31.04. 
4C\,31.04 is a low-redshift ($z = 0.0602$) CSS source with highly compact ($\sim 100$\,pc across) lobes believed to be $\sim 10^3$\,yr old\,\citep[][henceforth referred to as \citetalias{Giroletti2003}]{Giroletti2003}. The mottled and asymmetric radio morphology suggests strong interactions between the jets and a dense ISM.
The proximity of 4C\,31.04 enables us to probe jet-ISM interactions at the necessary sub-kpc scales with adaptive optics (AO)-assisted observations on an 8-metre telescope; nonetheless, no previous optical or near-infrared (IR) observations have resolved the host galaxy down to scales comparable to the size of the radio lobes. 

With the aim of observing jet-induced AGN feedback in action, we observed 4C\,31.04 with the Near-infrared Integral Field Spectrograph (NIFS) and the ALTAIR adaptive optics (AO) system on the Gemini North telescope in September 2016.
In our NIFS observations, we probe both the warm molecular and ionized gas phases, both of which are important tracers of jet-ISM interactions. 
Many groups have carried out similar studies of young radio galaxies in the past (e.g., 3C\,326\,N\,\citep{Nesvadba2010,Nesvadba2011}, 4C\,12.50\,\citep{Morganti2013}, IC\,5063\,\citep{Tadhunter2014,Morganti2015}, NGC\,1052\,\citep{Morganti2005}, NGC\,1068\,\citep{Riffel2014}, NGC\,1275\,\citep{Scharwacter2013}, NGC\,4151\,\citep{StorchiBergmann2012} and PKS\,B1718-649\,\citep{Maccagni2016b}).
However, the 100\,pc-scale jets of 4C\,31.04 provide a rare opportunity to observe jet-ISM interactions in the very earliest stages of evolution. Moreover, 4C\,31.04 has a wealth of auxiliary data at multiple wavelengths, including milliarcsecond-resolution very long baseline interferometry (VLBI) imaging, that enable us to better constrain the properties of the jets and of the host galaxy.

% Road map
In Section\,\ref{sec:Previous observations of 4C31.04} we summarise the properties of 4C\,31.04 and its host galaxy gleaned from previous multi-wavelength studies. Section\,\ref{sec:Observations and data reduction} details our NIFS observations and data reduction methods. 
In Section\,\ref{sec: Results} we discuss the two distinct phases of the ISM we detect in our observations.
We discuss the interpretation of our observations in the context of AGN feedback in Section\,\ref{sec:Discussion} and summarise our findings in Section\,\ref{sec:Conclusion}.

For the remainder of this paper, we assume a cosmology with $H_0 = 69.6$ km s$^{-1}$ Mpc$^{-1}$, $\Omega_M = 0.286$ and $\Omega_{\rm vac} = 0.714$, implying a luminosity distance $D_L = 271.3$\,Mpc and spatial scale of 1.17\,kpc arcsec$^{-1}$ at the redshift $z = 0.0602$ of 4C\,31.04\,\citep{Wright2006}.

%%%%%%%%%%%%%%%%%%%%%%%%%%%%%%%%%%%%%%%%%%%%%%%%%%%%%%%%%%%%%%%%%%%%%%%%%%%%%%%%%%%%%%%%%%%%%%%%%%%%%%%%%%%%%%%%%%%%%%%%%%%%%%%%%%%%%%%%%%%%%%%%%%%%%%%%%%%%%%%%%%%%%%%%%%%%%%

\section{Previous observations of 4C\,31.04}\label{sec:Previous observations of 4C31.04}

\subsection{Host galaxy properties} 
The host galaxy of 4C\,31.04 is MCG\,5-4-18, a giant elliptical approximately 2 $H$-band magnitudes brighter than $L_*$ at a redshift $z=0.0602 \pm 0.0002$\,\citep[][henceforth \citetalias{GarciaBurillo2007}]{GarciaBurillo2007}. \citet{Willett2010} provide an upper limit of $M_{\rm BH} \leq 10^{8.16}\,\rm \rm M_\odot$ on the mass of the central black hole using the width of the [O\,\textsc{iv}]$_{25.4\,\rm \mu m}$ line. 

This host galaxy has a Seyfert 2-like optical spectrum consistent with a predominantly old, metal-rich stellar population\,\citep{SeroteRoos2004a,Goncalves2004}. Despite this, there is evidence for moderate levels of star formation (SF): 
\citet{Willett2010} detect polycyclic aromatic hydrocarbon (PAH) emission and silicate absorption features in spatially unresolved \textit{Spitzer} mid-IR spectroscopy, indicating gas and dust heating by both ongoing SF and AGN. Using \textit{Infrared Astronomical Satellite (IRAS)} 60\,$\mu$m and 100$\,\mu$m fluxes \citet{OcanaFlaquer2010} estimate a star formation rate ${\rm SFR_{FIR}} \approx 4.9\,\rm M_\odot$ yr$^{-1}$, comparable to that calculated by \citet{Willett2010} using PAH emission features (${\rm SFR_{PAH}} = 6.4\,\rm M_\odot\,\rm yr^{-1}$).

\citet[][henceforth \citetalias{Perlman2001}]{Perlman2001} observed 4C\,31.04 with the \textit{Hubble Space Telescope} (\textit{HST}) using the Wide Field Planetary Cam 2 (WFPC2) and NICMOS, revealing several obscuring dust features, including an edge-on circumnuclear disc with a pronounced warp. Mid-IR silicate absorption features indicate that the dust has a clumpy distribution\,\citep{Willett2010}.
These and other multi-wavelength studies indicate that for a low-$z$ radio host galaxy, 4C\,31.04 harbours an unusually massive ($10^9\,\rm M_\odot$) multi-phase circumnuclear disc. This is discussed further in Section\,\ref{subsec:The circumnuclear medium of 4C31.04}. 

4C\,31.04 is the dominant member of a small group. 
Its closest companion is a spiral galaxy at a projected distance of $\sim 20\,\rm kpc$. The companion has a redshift $z = 0.0548$\,\citep{vandenBergh1970}, corresponding to a cosmological distance separation greater than $20\,\rm Mpc$ and a velocity offset of approximately $1560\,\rm km\,s^{-1}$. 
While it is possible that both galaxies are members of the same group, and that the apparent difference in redshift is due to their peculiar velocities, the velocity offset far exceeds the expected velocity dispersion for a small group ($\sim 200 - 400\,\rm km\,s^{-1}$). We therefore conclude that the companion is not a group member, and that a past interaction between the two is highly unlikely. 
Indeed, 4C\,31.04 does not show any sign of recent interaction such as tidal tails. On this basis, \citetalias{Perlman2001} conclude that any interaction must have taken place $\gtrsim 10^8$\,yr ago.

%%%%%%%%%%%%%%%%%%%%%%%%%%%%%%%%%%%%%%%%%%%%%%%%%%%%%%%%%%%%%%%%%%%%%%%%%%%%%%%%%%%%%%%%%%%%%%%

\subsection{Radio properties}
4C\,31.04 is a Compact Steep Spectrum (CSS) source, with $P_{\text{1.4 GHz}}=10^{26.3}$\,W\,Hz$^{-1}$\,\citep{vanBreugel1984a} and a spectral peak at $\sim 400$ MHz (estimated from the spectrum provided in the NASA/IPAC Extragalactic Database (NED)).
The source has two edge-brightened radio lobes separated by $\sim100$\,pc with a weak inverted-spectrum core\,(\citet{Cotton1995},\,\citetalias{Giroletti2003},\,\citet{Struve2012}). The axis of the jets is approximately East-West, with a $\rm PA \approx 100^\circ$.
%\followup{Could this be evidence of a dense ISM? AW says it could also be consistent with a puffy, turbulent ring.}
The lobes are highly asymmetric, suggesting strong jet-ISM interaction\,\citep{Giovannini2001}. The Western lobe is relatively faint and diffuse, suggesting the jet is interacting with many small clouds\,\citep{Bicknell2002}, whilst the Eastern lobe is brighter and more compact, and has a peculiar `hole' that may be caused by an overdensity in the ISM that is impenetrable to radio plasma\,\citepalias{Giroletti2003}. 

Radio observations indicate 4C\,31.04 is a truly young radio source.
Low flux variability ($\lesssim 2\rm\,per\,cent$ at 5\,GHz) and polarization ($\lesssim 1\rm\,per\,cent$ at 5\,GHz) in the lobes suggest beamed emission is improbable\,\citepalias{Giroletti2003}, and that the compactness of the source is unlikely an orientation effect.
\citet{Giovannini2001} estimated that the jets are nearly coplanar with the sky, with an orientation angle $\theta \gtrsim 75^\circ$. 
Differential VLBI imaging of 4C\,31.04 has revealed the radio emission to be rapidly expanding, with hotspot velocities of $\sim 0.3c$, yielding a dynamic age of $\sim 550$\,yr; synchrotron decay modelling yields much older radiative ages of $3300\,\rm yr$ and $4500-4900\,\rm yr$ in the Eastern and Western lobes respectively\,\citepalias{Giroletti2003}.
The difference in these estimates may be the result of the jet termini moving from point to point as the source evolves\,\citep{SutherlandBicknell2007}.
%\citetalias{SutherlandBicknell2007} pointed out that the discrepancy between the radiative and dynamic ages is consistent with 4C\,31.04 if it is at the end of the energy-driven bubble stage of evolution, which has lasted 3-5000\,yr, and that the 100\,pc-scale lobes have developed more recently ($< 1000$\,yr).

%%%%%%%%%%%%%%%%%%%%%%%%%%%%%%%%%%%%%%%%%%%%%%%%%%%
\subsection{Previous observations of the circumnuclear medium}\label{subsec:The circumnuclear medium of 4C31.04}

A number of multi-wavelength studies have shown that 4C\,31.04 has a multiphase and dynamically unrelaxed circumnuclear disc that contains dust, molecular gas and atomic gas.

The $R-H$ colour map of \citetalias{Perlman2001} constructed from $HST$ WFPC2/F702W and NICMOS/F160W images (our Fig.~\ref{fig: eline fluxes over R-H image} and their Fig.~2, bottom right) reveals several obscuring structures surrounding the nucleus, including a reddened circumnuclear disc, loops streaming from the Northernmost point of the disc, and a large S-shaped structure extending to the North and South. 
The circumnuclear disc is approximately perpendicular to the axis of the radio jets, and extends roughly 500\,pc to the North and 1000\,pc to the South. The disc is highly inclined, and is viewed almost edge-on. 

Using the Institut de Radioastronomie Millim\'etrique (IRAM)\,30\,m telescope, \citet{OcanaFlaquer2010} found double-horned $^{12}$CO(2--1) and $^{12}$CO(1--0) profiles with superimposed absorption, suggesting a massive inclined disc of molecular gas and a total \hh{} mass of $(60.63 \pm 16.92) \times 10^8 \,\rm M_\odot$.
\citetalias{GarciaBurillo2007} also detected a disc-like structure in the 1\,mm continuum with the IRAM Plateau de Bure interferometer (PdBI) $\sim 1.4$\,kpc in size which is consistent with the dusty disc in the \textit{HST} $R-H$ image. They also found \hco{} $(1-0)$ emission to the North and South of the nucleus, with kinematics consistent with a disc (their Fig.~1b). 
\citet{Struve2012} detected \hi{} absorption using VLBI observations, and estimated a column density $N_\text{H\,\textsc{i}} = 1.2-2.4\times10^{21}$\,cm$^{-2}$.
The velocity structure of the absorption is consistent with a large rotating disc of atomic gas coinciding with the kpc scale molecular disc.
The \hi{} optical depth is much higher in the Eastern lobe, indicating that the disc is inclined such that the Eastern lobe is viewed through a dense column of gas. This is consistent with earlier \hi{} VLBI observations by \citet{Conway1996}, which also revealed a sharp `edge' in the \hi{} opacity in the Western lobe (their Fig.~1).
\citet{Conway1996} also detected high-velocity \hi{} clouds and free-free absorption features in front of both lobes, which they attributed to the jets evaporating material off the inner edge of the circumnuclear disc. 

Irregularities in the kinematics and morphology of gas and dust in the circumnuclear regions of the galaxy suggest unrelaxed dynamics.
\citetalias{Perlman2001} fitted isophotes to the \textit{HST} NICMOS F160W ($H$-band) image and found significant anisotropies in the ellipticity and a position angle (PA) that twists by tens of degrees in the innermost 2--4\,kpc. 
\citet{Struve2012} detected a narrow, redshifted H\,\textsc{i} absorption component consistent with a cloud at a radius $\gg 100$\,pc falling into the nucleus, which may be a remnant of a merger or accretion event.
\citetalias{GarciaBurillo2007} detected blueshifted ($\sim 150$\,km\,s$^{-1}$) \hco{} $(1-0)$ absorption over the centre of the galaxy, although we note that the uncertainty in the redshift of the source corresponds to an uncertainty in the galaxy's systemic velocity of $\sim 100\,\rm km\,s^{-1}$.

% Summary sentence for this section
These observations show that the host of 4C\,31.04 contains a dense, massive, circumnuclear disc, consisting of cold dust, both cold and warm molecular gas, and atomic gas. 
Twisted central isophotes and non-circular motions indicate that the disc is dynamically unrelaxed, perhaps due to a previous a merger or interaction, accretion of new material, or by jet-ISM interactions.

%%%%%%%%%%%%%%%%%%%%%%%%%%%%%%%%%%%%%%%%%%%%%%%%%%%
\section{Observations and data reduction}\label{sec:Observations and data reduction}

\subsection{Observations}
We observed 4C\,31.04 using the Near-infrared Integral Field Spectrograph (NIFS)\,\citep{McGregor2003} on the 8.1-m Gemini North telescope on Mauna Kea in Hawai`i. 
NIFS provides $R\sim 3500$ $J$-, $H$- and $K$-band spectroscopy over a 3'' $\times$ 3'' field of view with 0.1'' $\times$ 0.04'' spaxels. 
NIFS is fed by the ALTAIR adaptive optics (AO) system which can be used in laser guide star (LGS) or natural guide star (NGS) mode to provide near-diffraction limited resolution.

We obtained $H$- and $K$-band observations of 4C\,31.04 using NIFS and the ALTAIR AO system used in LGS mode, using an off-axis guide star for tip/tilt correction, with a position angle of 0$^\circ$ on 2016 September 22 during program GN-2016B-C-1. 
We used 600 s exposure times for both source and sky frames, integrating on-source for a total of 80 and 60 minutes in the $H$- and $K$-bands respectively. 
The HIPPARCOS stars HIP12218, HIP117774 and HIP12719, observed before and after 4C\,31.04, were used as telluric and flux standards. 
The angular resolution (Full Width at Half Maximum (FWHM)) of our observations was approximately $0.20''$ and $0.17''$ in the $H$- and $K$-bands respectively, which we estimated by fitting Gaussian profiles to our standard star exposures.

\subsection{NIFS data reduction in \textsc{IRAF}} 
We reduced the data using the \textsc{Gemini IRAF} package, reducing science frames for both object and standard star exposures as follows.

We subtract individual sky frames from the science frames, pairing the sky frames taken closest in time to the science frame.
We divide by a master flat field, then extract the slices from the science frames to form 3D data cubes. All data cubes are spatially interpolated to yield 0.05'' $\times$ 0.05'' square pixels.
We then interpolate over bad pixels, apply the wavelength calibration and correct for spatial distortion. The wavelength solution was found using argon and xenon arc lamp exposures taken during the night, and spatial distortions were calibrated using an exposure of a Ronchi grating through a flat field.

% Telluric
To correct for telluric absorption lines, we generate a 1D spectrum of the telluric standard by co-adding the spectra within an 0.5''-diameter aperture centred on the star. We flatten the resulting 1D spectrum by dividing it by a normalised blackbody corresponding to the temperature of the star, estimated from its spectral class, and then fit and remove stellar absorption lines using a Voigt profile.
We remove telluric absorption lines by dividing the object data cubes by the resulting 1D spectrum.

% Flux calibration
To flux calibrate the object data cubes, we use exposures of a standard star with given 2MASS $H$-band and $Ks$-band magnitudes to convert counts to units of monochromatic flux density. 
We generate a 1D spectrum of the flux standard in the same fashion as for the telluric standard, and multiply it by the normalised blackbody to restore its original spectral shape.
We remove telluric absorption lines by dividing the spectrum by the 1D spectrum of the telluric standard.
We then divide it by a blackbody spectrum in units of $\rm erg\,s^{-1}\,cm^{-2}$\,\AA$^{-1}$ corresponding to the temperature of the flux calibration standard star.  
We fit a polynomial to the resulting spectrum to give the transfer function. 
To flux calibrate the object data cubes, we divide the individual object data cubes by the transfer function, the exposure time and the spaxel area to give units of $\rm erg\,s^{-1}\,cm^{-2}$\,\AA$\rm ^{-1}\,arcsec^{-2}$. 
Finally, we shift and median-combine each object data cube to yield a single data cube.

\subsection{MAD smoothing} We use a Median Absolute Deviation (MAD) smoothing algorithm with a radius of 2 pixels to smooth the reduced data cubes and to remove artefacts. 
In each wavelength slice, for each pixel, we compute the median and standard deviation $\sigma$ of the surrounding pixels out to the specified radius, and reject those pixels with absolute value greater than $3\sigma$ from the median. We iterate until no more pixels are rejected. The value of the central pixel is then replaced by the mean of the remaining pixels, and the variance of the central pixel is replaced by the mean of the variance of the remaining pixels.

Before smoothing, the angular resolution of our observations corresponds to a FWHM of approximately 230\,pc and 200\,pc in the $H$- and $K$-bands respectively. 
After smoothing with a radius of 2 pixels, the resolution is decreased to approximately 300\,pc and 280\,pc in the $H$- and $K$-bands respectively. 

\subsection{Emission line fitting}
We use \textsc{mpfit}\,\citep{Markwardt2009}, a \textsc{python} implementation of the Levenberg-Marquardt algorithm\,\citep{More1978} developed by M. Rivers\footnote{Available \url{http://cars9.uchicago.edu/software/python/mpfit.html}.} to fit single-component Gaussian profiles to emission lines. 
We keep fits with $\chi^2 < 2$ and signal-to-noise ratio $\rm(SNR) > 1$. In all reported linewidths, we have accounted for instrumental resolution by subtracting the width of the line spread function (LSF) in quadrature from the width of the fitted Gaussian. We estimate the width of the LSF by fitting a Gaussian to sky lines close in wavelength to the relevant emission line.

Integrated line fluxes and upper limits for emission lines in 4C\,31.04 are shown in Table~\ref{tab:measured line fluxes}.
To calculate integrated line fluxes, we simply sum the fluxes found in each spaxel.

To calculate upper limits for line fluxes which are not detected using our $\chi^2$ and SNR criteria, we use the following method.
In each spaxel, we calculate the standard deviation $\sigma$ in the continuum in a window centred on the emission line. 
We assume the non-detected emission line in that spaxel is a Gaussian with amplitude $3\sigma$. 
The width we use depends on the emission line. For non-detected ro-vibrational \hh{} lines, we use the measured width of the \hh{}\,1--0\,S(1) emission line in that spaxel. 
For hydrogen recombination lines, we use the Gaussian sigma of 24.9\,\AA{} we calculate from the measured equivalent width of 18\,\AA{} for the combined H$\alpha$ and [N\,\textsc{ii}] lines in a single-slit optical spectrum of 4C\,31.04 reported by \citet{Marcha1996}. When quoting integrated upper limits, we assume the lines are detected in every spaxel in which we detect the \hh{}\,1--0\,S(1) emission line.

\begin{table*}
	\centering
	\caption{Emission line fluxes and their uncertainties. The integrated flux is measured by adding together the emission line fluxes in each individually fitted spaxel. Upper limits are computed using the method described in Section\,\ref{sec:Observations and data reduction}. All quantities are given in units of erg s$^{-1}$ cm$^{-2}$.
}
	\label{tab:measured line fluxes}
	\begin{tabular}{c c c c}
	\hline
	\textbf{Emission line} & \textbf{Total Integrated flux} ($\rm erg\,s\,cm^{-2}$) & \textbf{Integrated flux within} $r \leq 0.2''$ & \textbf{Integrated flux outside} $r > 0.2'' $ \\
	\hline 
	\hh{}\,1--0\,S(1) & $2.60 \pm 0.06 \times 10^{-15}$ & $ 5.2 \pm 0.2 \times 10^{-16}$ & $2.08 \pm 0.04 \times 10^{-15}$ \\
	\hh{}\,1--0\,S(2) & $4.4 \pm 0.5 \times 10^{-16}$   & $ 2.0 \pm 0.2 \times 10^{-16}$ & $2.3 \pm 0.2 \times 10^{-16}$ \\
	\hh{}\,1--0\,S(3) & $2.52 \pm 0.11 \times 10^{-15}$ & $ 6.3 \pm 0.3 \times 10^{-16}$ & $1.90 \pm 0.05 \times 10^{-15}$ \\
	\hh{}\,2--1\,S(1) & $\leq 3.077 \times 10^{-15}$ & $\leq 2.009 \times 10^{-16}$ & $\leq 2.876 \times 10^{-15}$ \\
	\hh{}\,2--1\,S(2) & $\leq 2.172 \times 10^{-15}$ & $\leq 1.667 \times 10^{-16}$ & $\leq 2.005 \times 10^{-15}$ \\
	\hh{}\,2--1\,S(3) & $\leq 3.222 \times 10^{-15}$ & $\leq 1.972 \times 10^{-16}$ & $\leq 3.025 \times 10^{-15}$ \\
	\hi{} Br\,$\gamma$ & $\leq 4.500 \times 10^{-15}$ & $\leq 3.411 \times 10^{-16}$ & $\leq 4.159 \times 10^{-15}$ \\
	\feiilong{} & $1.35 \pm 0.05 \times 10^{-15}$ & --- & --- \\	
	\hline	
	\end{tabular}
\end{table*}

\begin{table}
	\centering
	\caption{Ratios of the integrated fluxes of emission lines evaluated over the entire field of view (left column), within the central 0.2'' (middle column) and outside the central 0.2'' (right column). The integrated flux is measured by adding together the emission line fluxes in each individually fitted spaxel. Upper limits are computed using the method described in Section\,\ref{sec:Observations and data reduction}. 
	}
	\label{tab: line ratios}
	\begin{tabular}{c l l l}
		\hline
		\textbf{Emission line ratio} & \textbf{Total} & \textbf{Within $r \leq 0.2''$} & \textbf{Outside $r > 0.2'' $} \\
		\hline 
		\hh{}\,1--0\,S(1)/\hi{} Br\,$\gamma$ & 0.5774 & 1.5212 & 0.5000 \\
		\hh{}\,1--0/2--1\,S(1) & 0.8443 & 2.5826 & 0.7229 \\
%		\feiilong{}/Br\,$\gamma$ & 1.06294 & 1.06294 & --- \\
		\hline
	\end{tabular}
\end{table}

\section{Results}
\label{sec: Results}

\begin{figure*}
	\centering
	\subcaptionbox{\label{fig: K-band continuum over B image}}{		
		\includegraphics[width=0.48\textwidth]{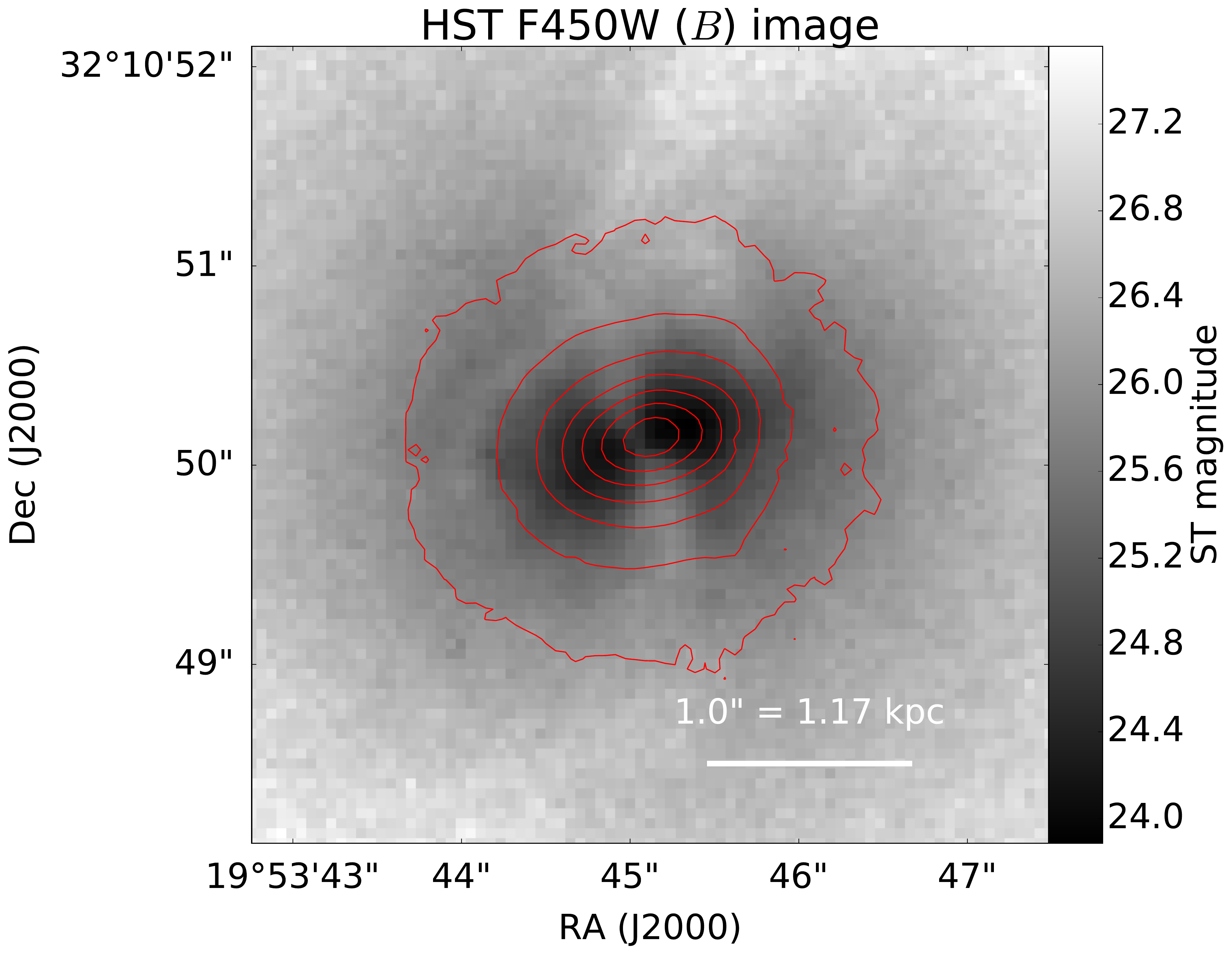}
	}
	\hfill
	\subcaptionbox{\label{fig: eline fluxes over R-H image}}{		
		\includegraphics[width=0.48\textwidth]{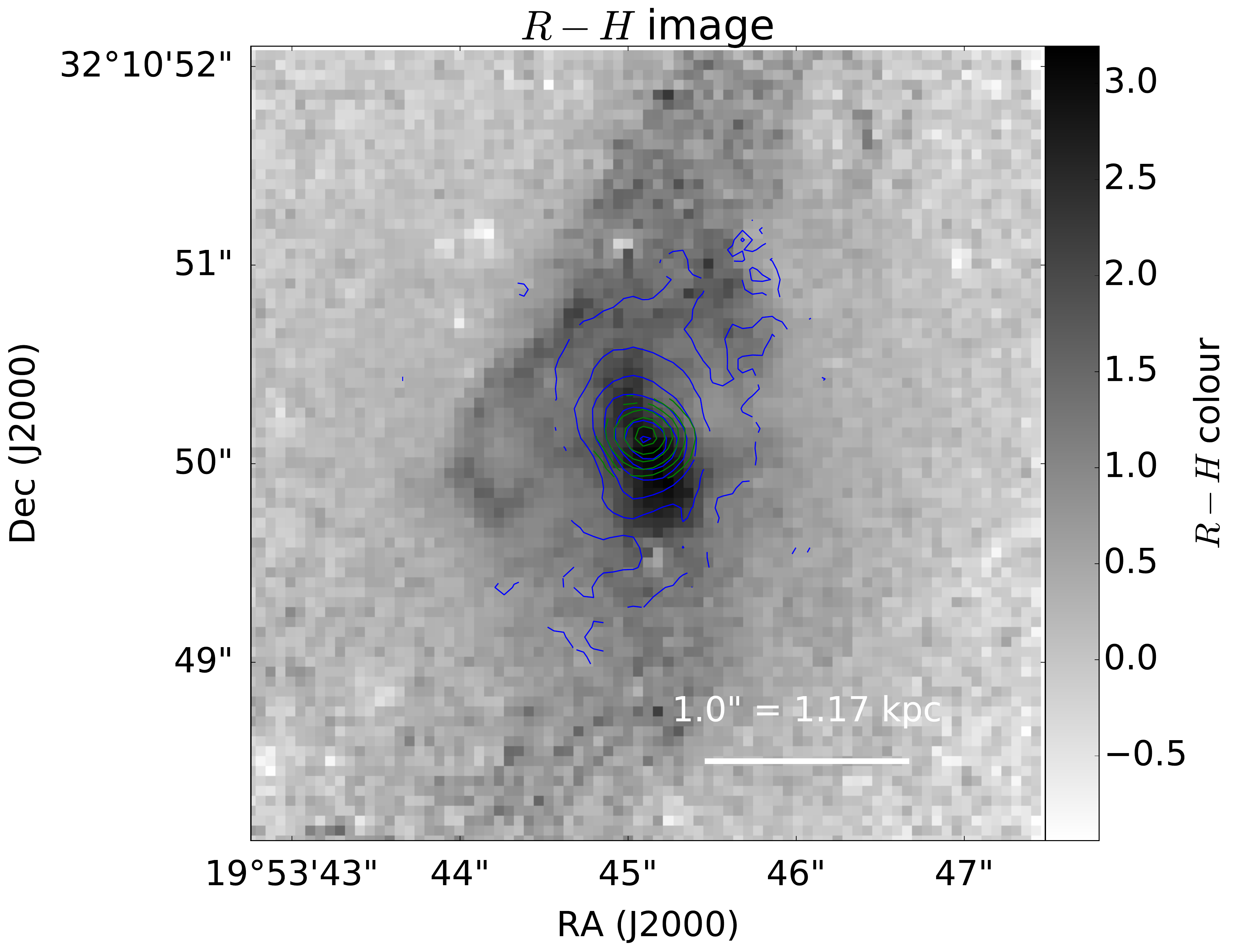}
	}
	\caption{
		Our $K$-band NIFS observations overlaid on to \textit{Hubble Space Telescope} images.
		(a) shows the NIFS $K$-band continuum (red contours) overlaid on to the \textit{HST} WFPC2 $B$-band (F450W) image. The $K$-band contours represent the underlying stellar mass distribution, showing that the `cones' to the East and West of the nucleus are not physical features, but a result of dust obscuration.
		The contours in (b) show the flux of the \hh{}\,1--0\,S(1) (blue) and \feiilong{} (green) emission lines overlaid on to the $R-H$ image constructed using the $HST$ WFPC2/F702W and NICMOS/F160W images. Whilst the \feii{} emission is concentrated to the central few 100\,pc, the \hh{} emission extends to $\sim 1\,\rm kpc$ North and South of the nucleus, suggesting it is part of the massive circumnuclear disc.}
	\label{fig: HST overlays}
\end{figure*}

Diffraction-limited \textit{HST} WFPC2 imaging has similar angular resolution to our NIFS observations ($0.05$''), enabling us to directly compare the two sets of observations. 
Fig.~\ref{fig: K-band continuum over B image} shows the $K$-band continuum (red) from our NIFS observations overlaid on to the \textit{HST} $B$-band image. Fig.~\ref{fig: eline fluxes over R-H image} overlays the fluxes of the ro-vibrational \hh{} emission (blue), which we detect in the $K$-band, and the \feiilong{} emission (green), which we detect in the $H$-band, on to the \textit{HST} $R-H$ image, placing both emission lines in context. The \hh{} emission traces the dusty disc, and shows hints of a warp to the North and South, suggesting the \hh{} is part of the large S-shaped dust feature. Meanwhile, the \feii{} emission is localised to the nucleus.

%%%%%%%%%%%%%%%%%%%%%%%%%%%%%%%%%%%%%%%%%%%%%%%%%%%
\subsection{Nuclear [Fe\,\textsc{ii}] emission}\label{sec:Nuclear Fe emission}

\begin{figure*}
	\subcaptionbox{\label{fig:[Fe II] continuum}}{		
	\includegraphics[width=0.48\textwidth]{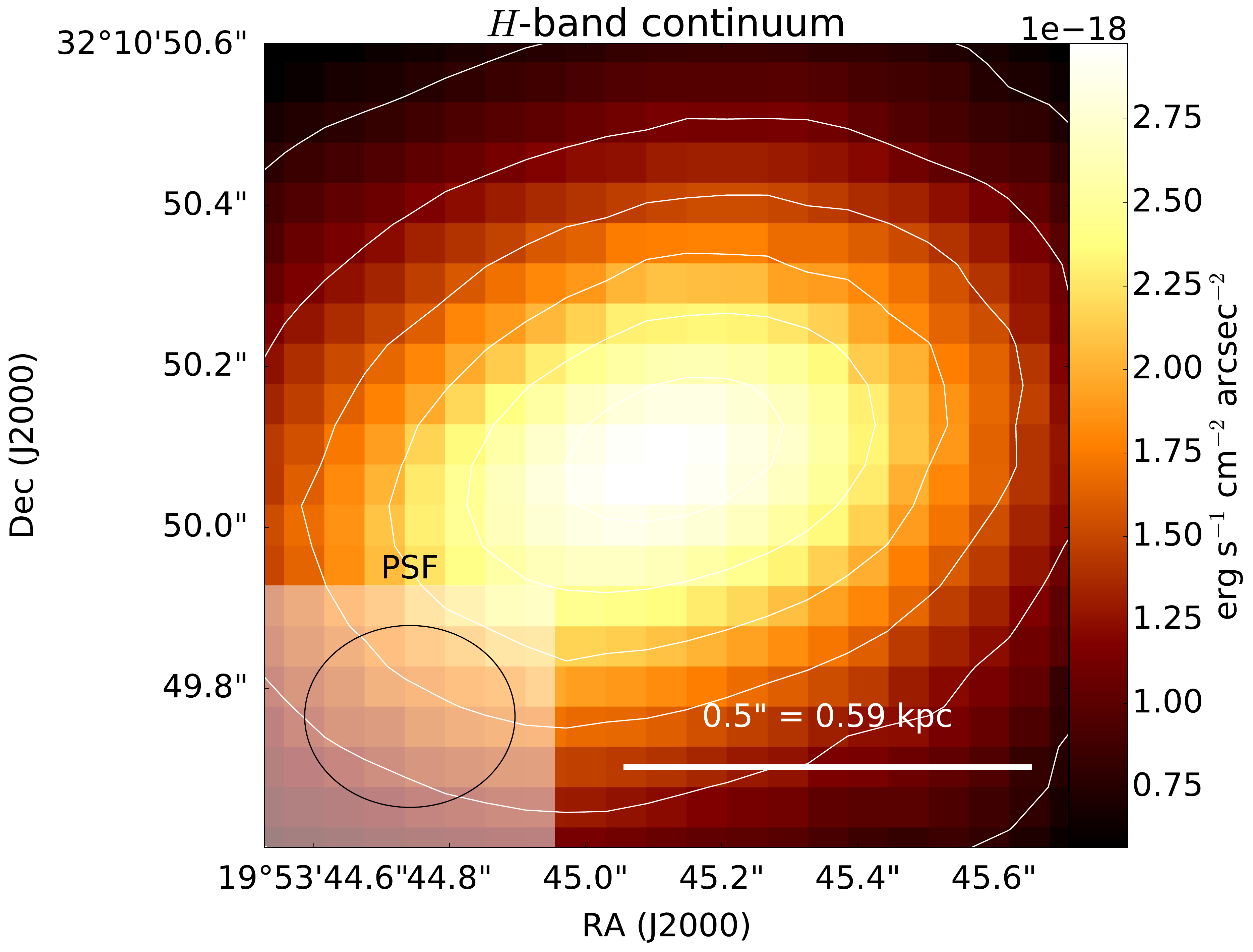}
	}
	\subcaptionbox{\label{fig:[Fe II] flux}}{		
		\includegraphics[width=0.48\textwidth]{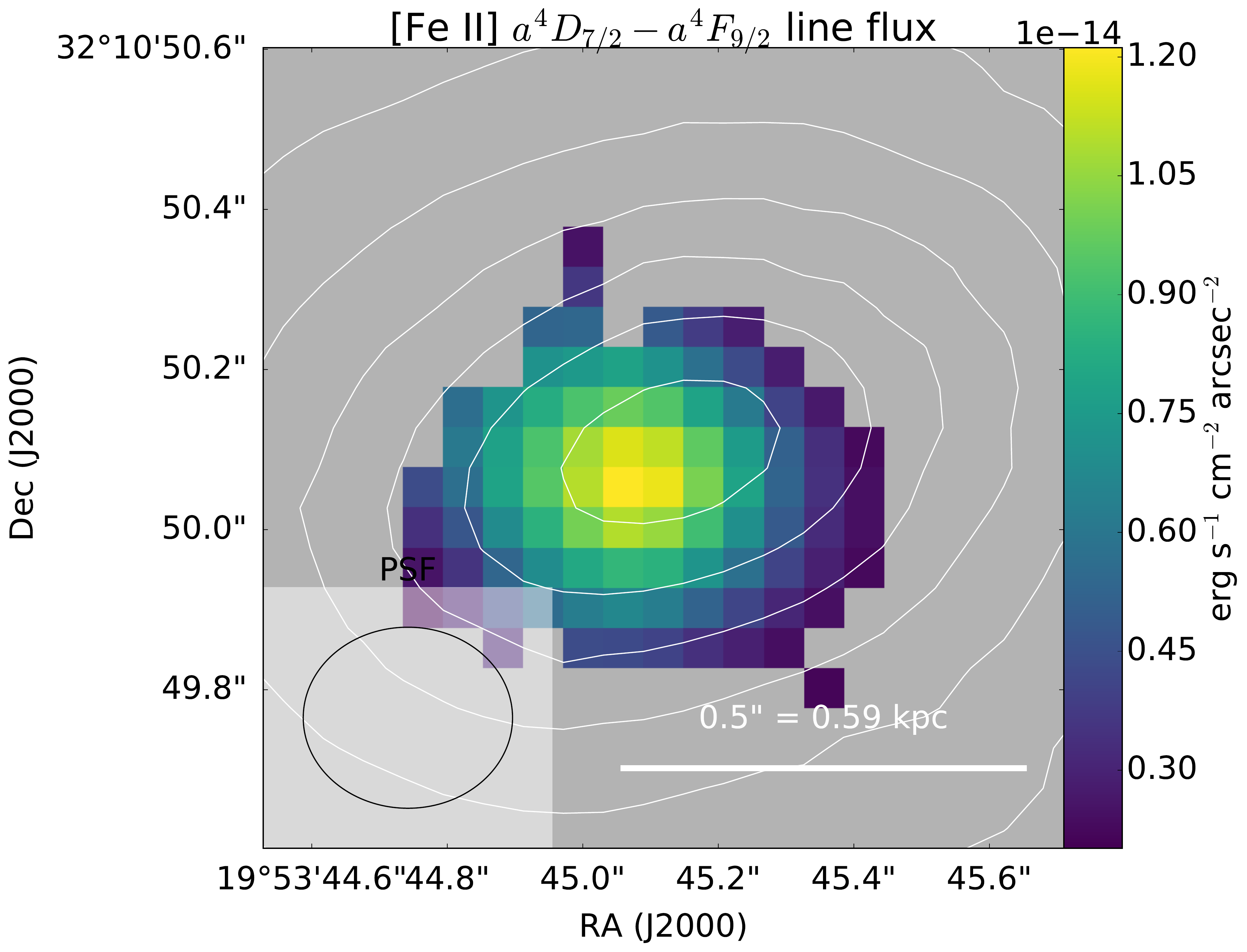}
	}
	\subcaptionbox{\label{fig:[Fe II] vrad}}{		
		\includegraphics[width=0.48\textwidth]{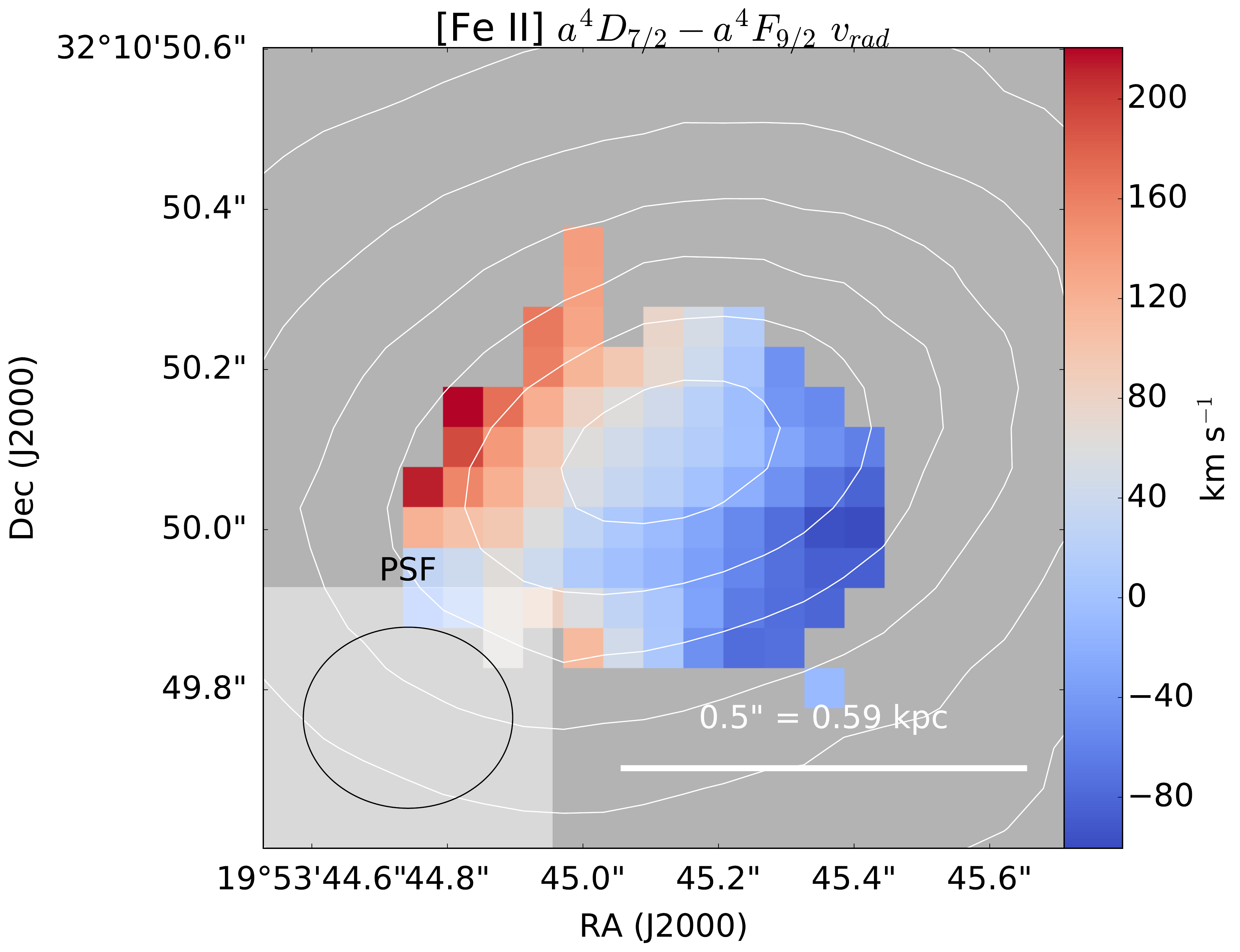}
	}
	\subcaptionbox{\label{fig:[Fe II] sigma}}{		
		\includegraphics[width=0.48\textwidth]{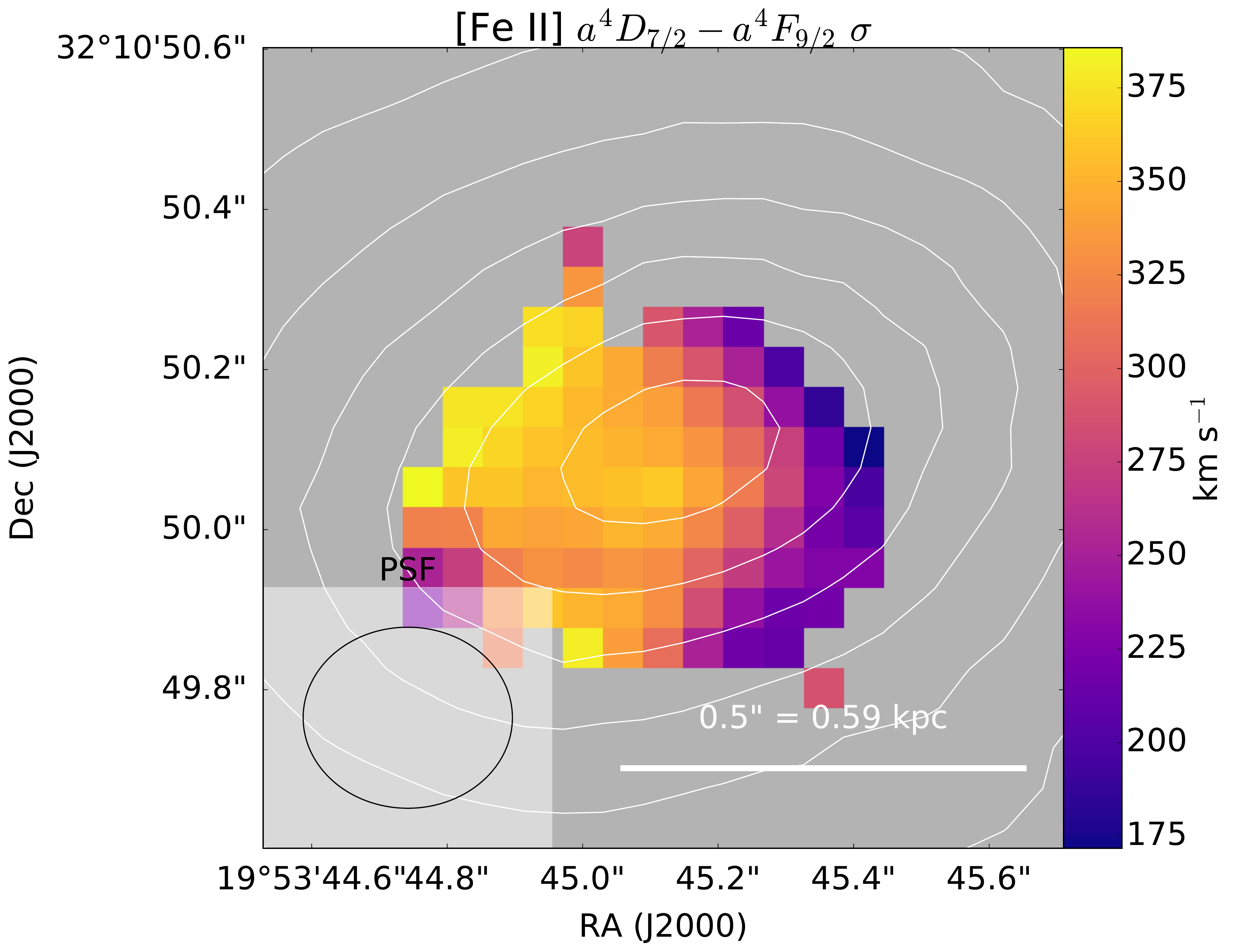}
	}
	\caption{
		(a): the $H$-band continuum; (b): \feiilong{} integrated flux; (c): \feiilong{} radial velocity (minus the systemic velocity of the galaxy obtained from the redshift of $z = 0.0602$); and (d): \feiilong{} velocity dispersion (Gaussian $\sigma$). The $H$-band continuum is indicated in contours and the FWHM of the PSF (taking into account the effects of MAD smoothing) is indicated in all figures.
	}
	\label{fig:[Fe II] maps}
\end{figure*}

In our $H$-band observations we detect [Fe\,\textsc{ii}] $a^4D_{7/2}-a^4F_{9/2}$ (rest-frame wavelength 1.644\,$\mu$m) emission in the inner few $100\,\rm pc$ of 4C\,31.04. 
Fig.~\ref{fig:[Fe II] maps} shows the $H$-band continuum and the [Fe\,\textsc{ii}] line flux, radial velocity and velocity dispersion.

% spatial extent
We measure the spatial extent of the \feii{} emission by fitting a 2D Gaussian to the integrated flux map. 
The emitting region is marginally resolved in our observations and is elongated, extending over $\approx 380$\,pc E-W and $\approx 320$\,pc N-S. 
% kinematics
The line profile is flat, asymmetric and broad, with a velocity dispersion of approximately 350\,km\,s$^{-1}$ across the emitting region.
We argue that the \feii{} emission traces gas being accelerated out of the plane of the circumnuclear disc by the jet-driven bubble, which we discuss further in Section\,\ref{sec:Discussion}.

%%%%%%%%%%%%%%%%%%%%%%%%%%%%%%%%%%%%%%%%%%%%%%%%%%%
\subsection{\hh{} emission}\label{sec:H2 emission}

\begin{figure*}
	\subcaptionbox{\label{fig:H2 continuum}}{		
		\includegraphics[width=0.48\textwidth]{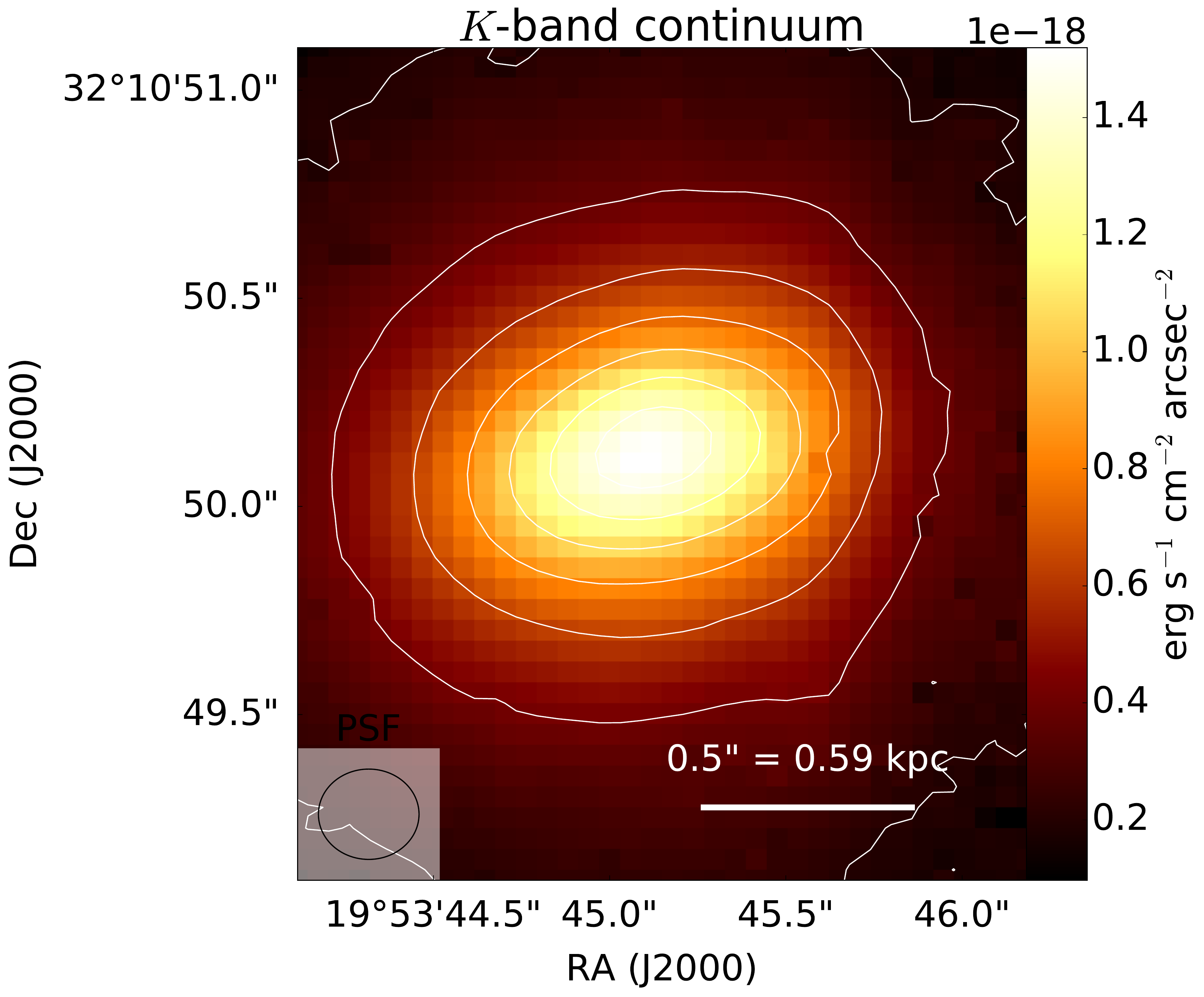}
	}
	\subcaptionbox{\label{fig:H2 flux}}{		
		\includegraphics[width=0.48\textwidth]{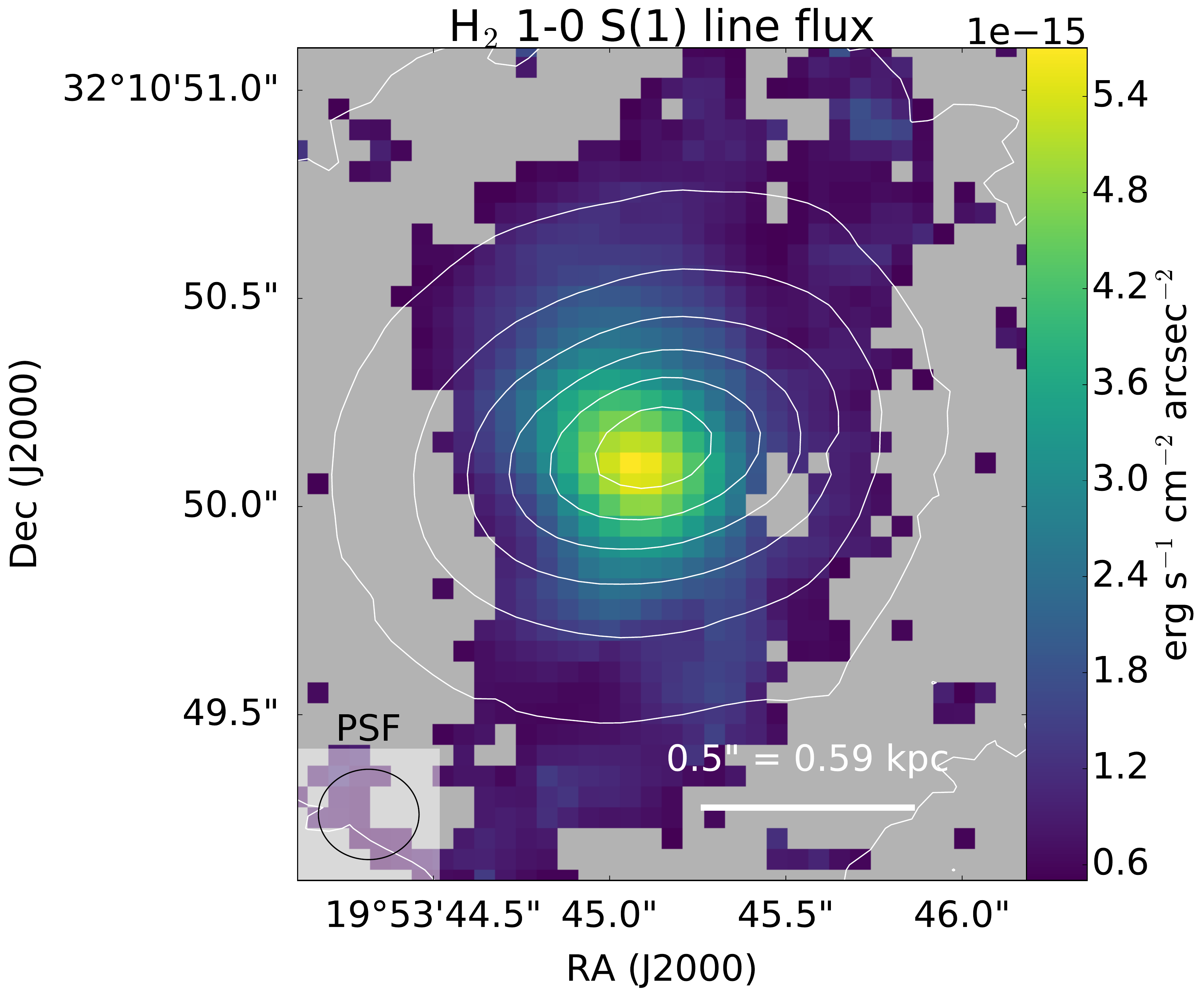}
	}
	\subcaptionbox{\label{fig:H2 vrad}}{		
		\includegraphics[width=0.48\textwidth]{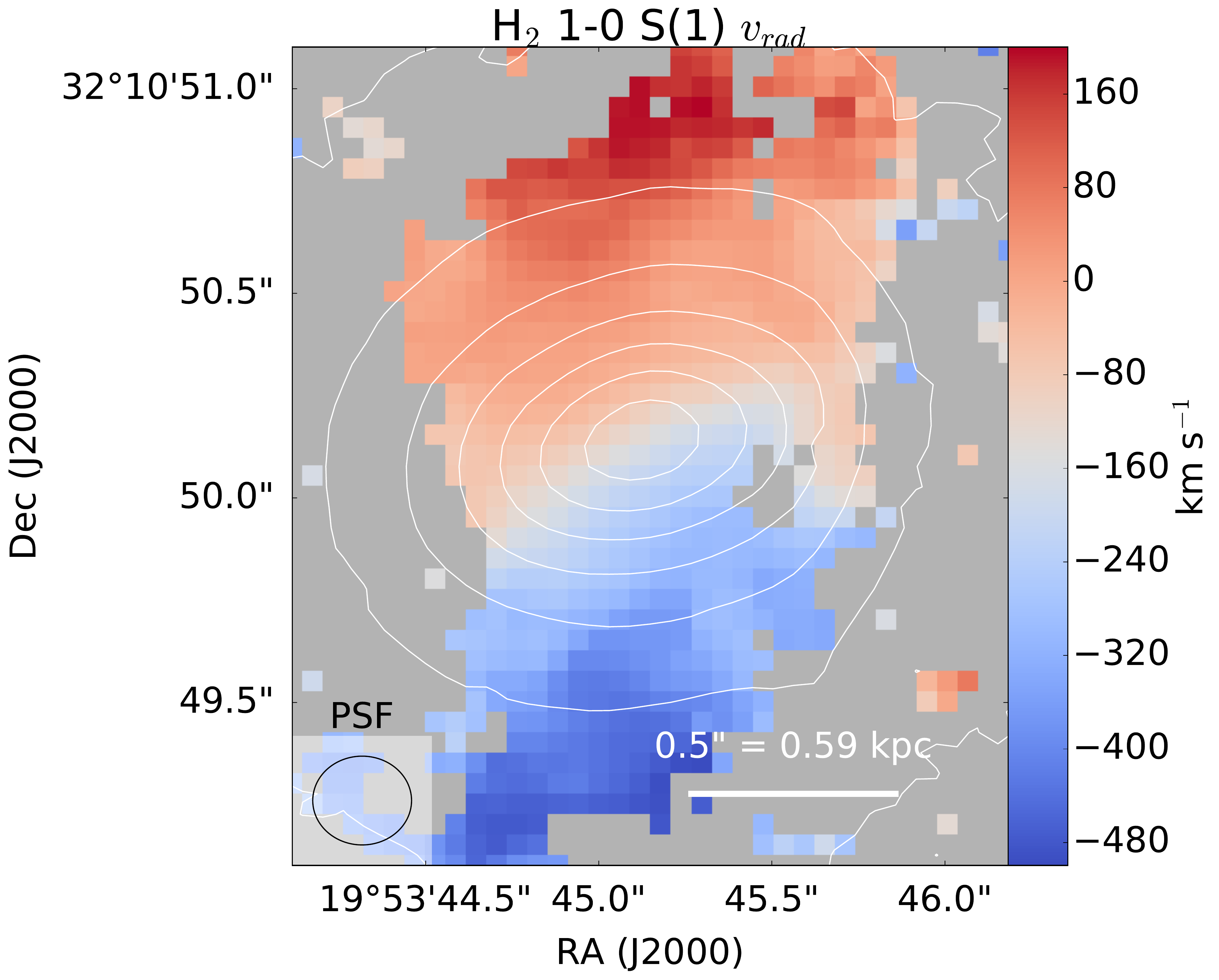}
	}
	\subcaptionbox{\label{fig:H2 sigma}}{		
		\includegraphics[width=0.48\textwidth]{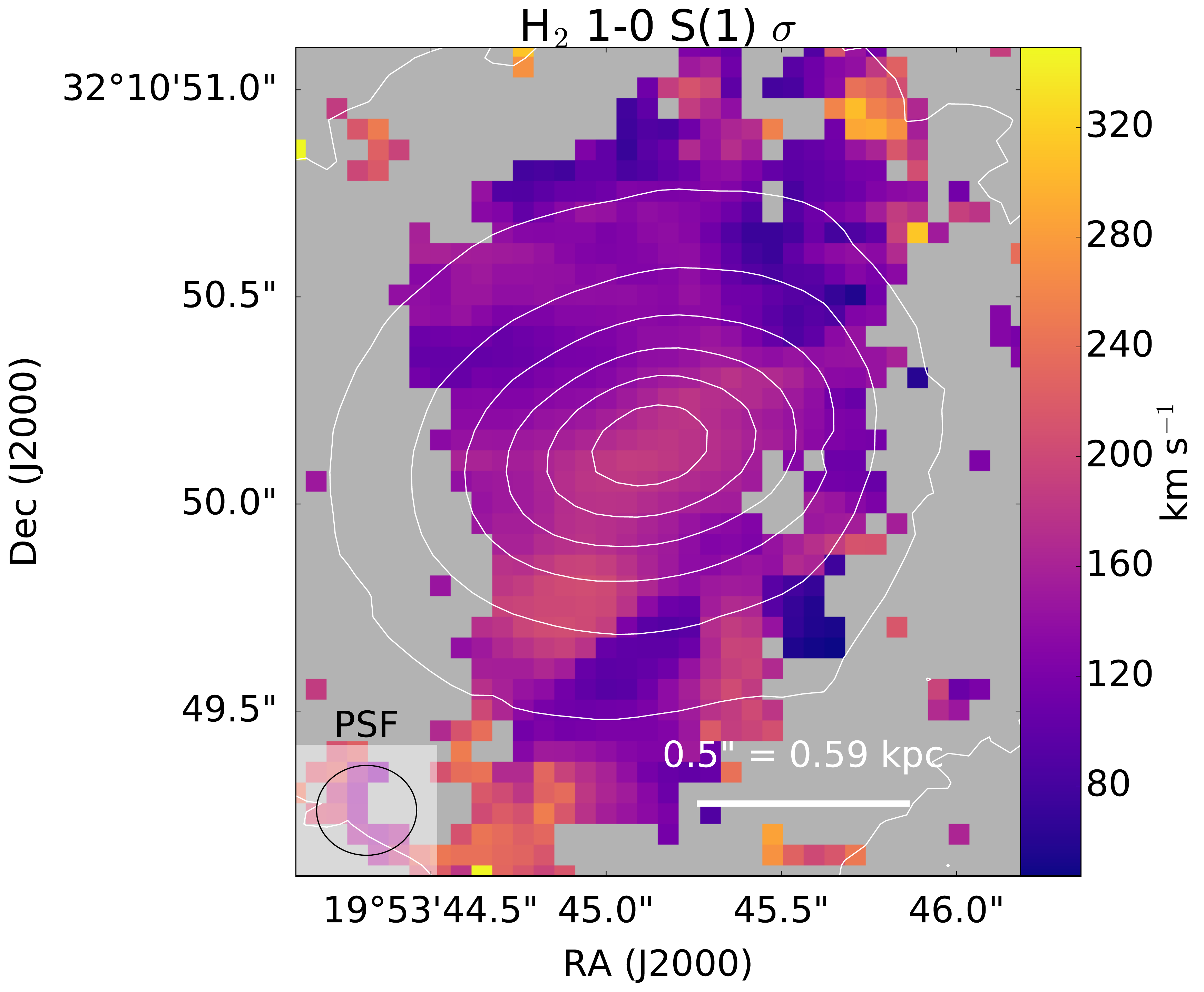}
	}
	\caption{
		 (a) the $K$-band continuum; (b): \hh{}\,1--0\,S(1) integrated flux; (c): \hh{}\,1--0\,S(1) radial velocity (minus the systemic velocity of the galaxy obtained from the redshift of $z = 0.0602$); and (d): \hh{}\,1--0\,S(1) velocity dispersion (Gaussian $\sigma$). The $K$-band continuum is indicated in contours and the FWHM of the PSF (taking into account the effects of MAD smoothing) is indicated in all figures.
	}
	\label{fig:H2 maps}
\end{figure*}

We detect the \hh{}\,1--0\,S(1), S(2) and S(3) emission lines in our $K$-band spectra, corresponding to transitions involving changes in the rotational quantum number with $\Delta J = +2$ and the vibrational level transition $\nu = 1$ to $\nu = 0$. These \textit{ro-vibrational} emission lines trace warm \hh{} in the temperature range $\sim 10^3-10^4$\,K. 
In Fig.~\ref{fig:H2 maps}, we show the flux distribution, radial velocity and velocity dispersion of the \hh{}\,1--0\,S(1) line we detect in our NIFS observations. 
The flux extends over $\approx 2 \rm\,kpc$ from North to South, and the radial velocity shows large-scale rotation, indicating that the warm \hh{} is a part of the kpc-scale circumnuclear disc observed in CO, \hco{} and \hi. 
The flux peaks sharply at the nucleus (Fig.~\ref{fig:H2 flux}): indeed, we find that $\approx 20\rm\,per\,cent$ of the total flux is contained within the central 0.4''. In this section, we analyse the inner 0.4'' separately to the remaining \hh{} emission in order to determine any differences in the excitation mechanism and temperature between the central and extended components.

\subsubsection{Kinematics}

\begin{figure*}
	\centering
	\subcaptionbox{\label{fig: disc fit measured}}{		
	\includegraphics[width=0.48\textwidth]{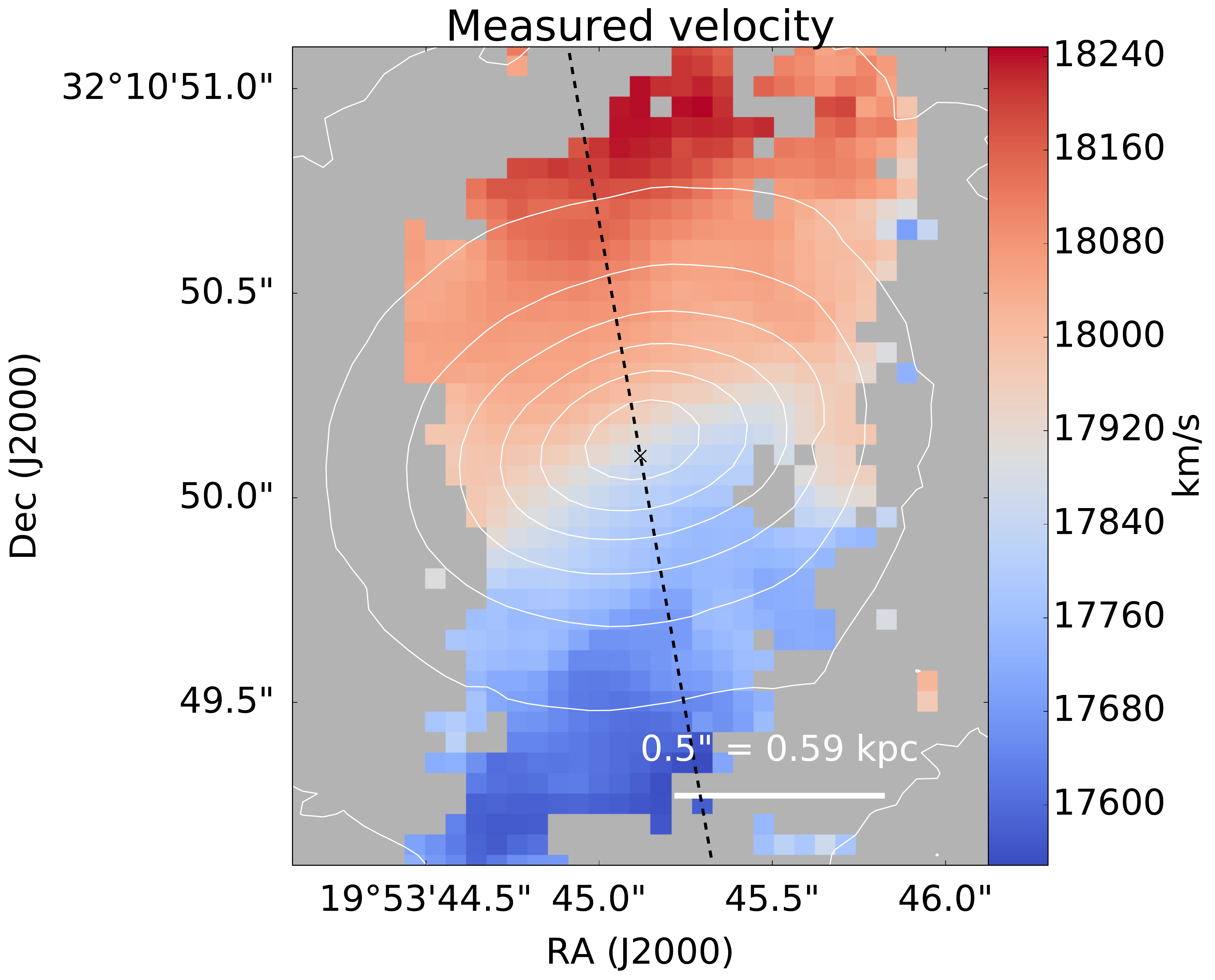}
	}
	\subcaptionbox{\label{fig: disc fit model}}{		
		\includegraphics[width=0.48\textwidth]{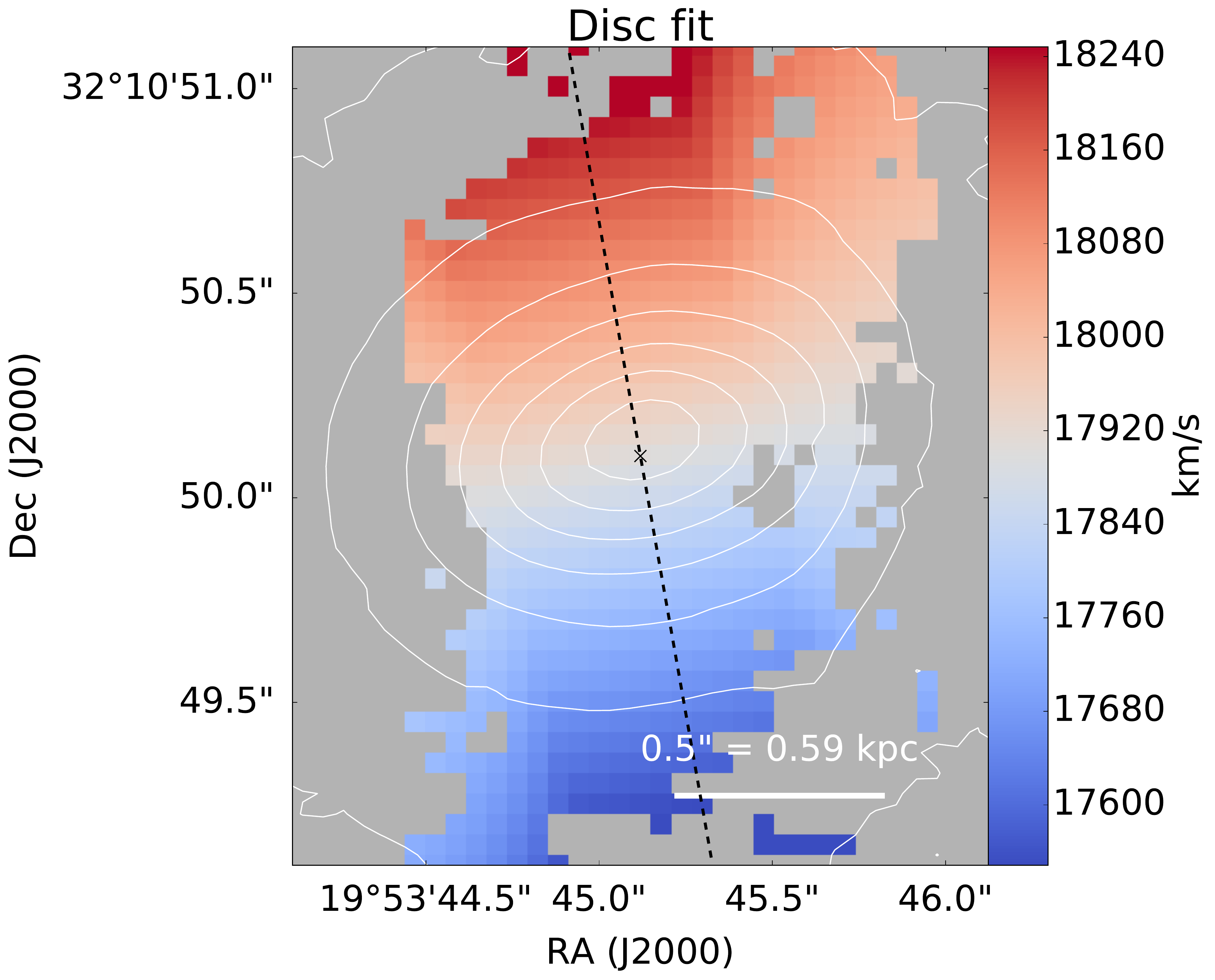}
	}
	\subcaptionbox{\label{fig: disc fit residual}}{		
		\includegraphics[width=0.48\textwidth]{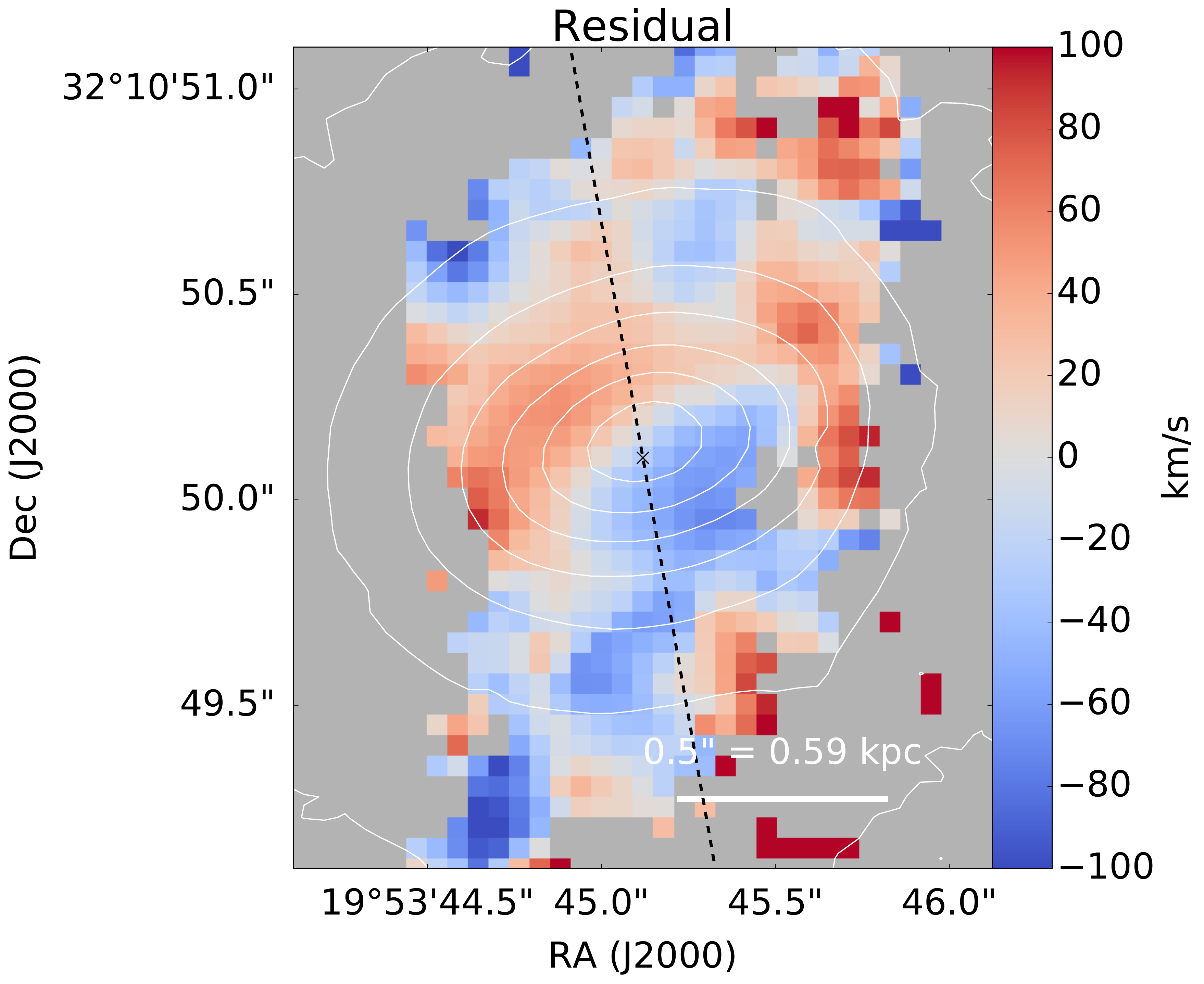}
	}
	\caption{Kinematics of the \hh{}\,1--0\,S(1) emission line; (a): measured radial velocity, (b) radial velocity of the model fit, and (c) the residual. With the exception of the residual plot, all velocities are given with respect to the local standard of rest. The cross represents the location of the nucleus (i.e., the peak in the $K$-band continuum). Fig.~\ref{fig: disc fit PV diagram} shows a cross-section of the radial velocities taken along the dashed black line.
	}
	\label{fig: disc fit}
\end{figure*}

\begin{figure}
	\centering
	\includegraphics[width=1\linewidth]{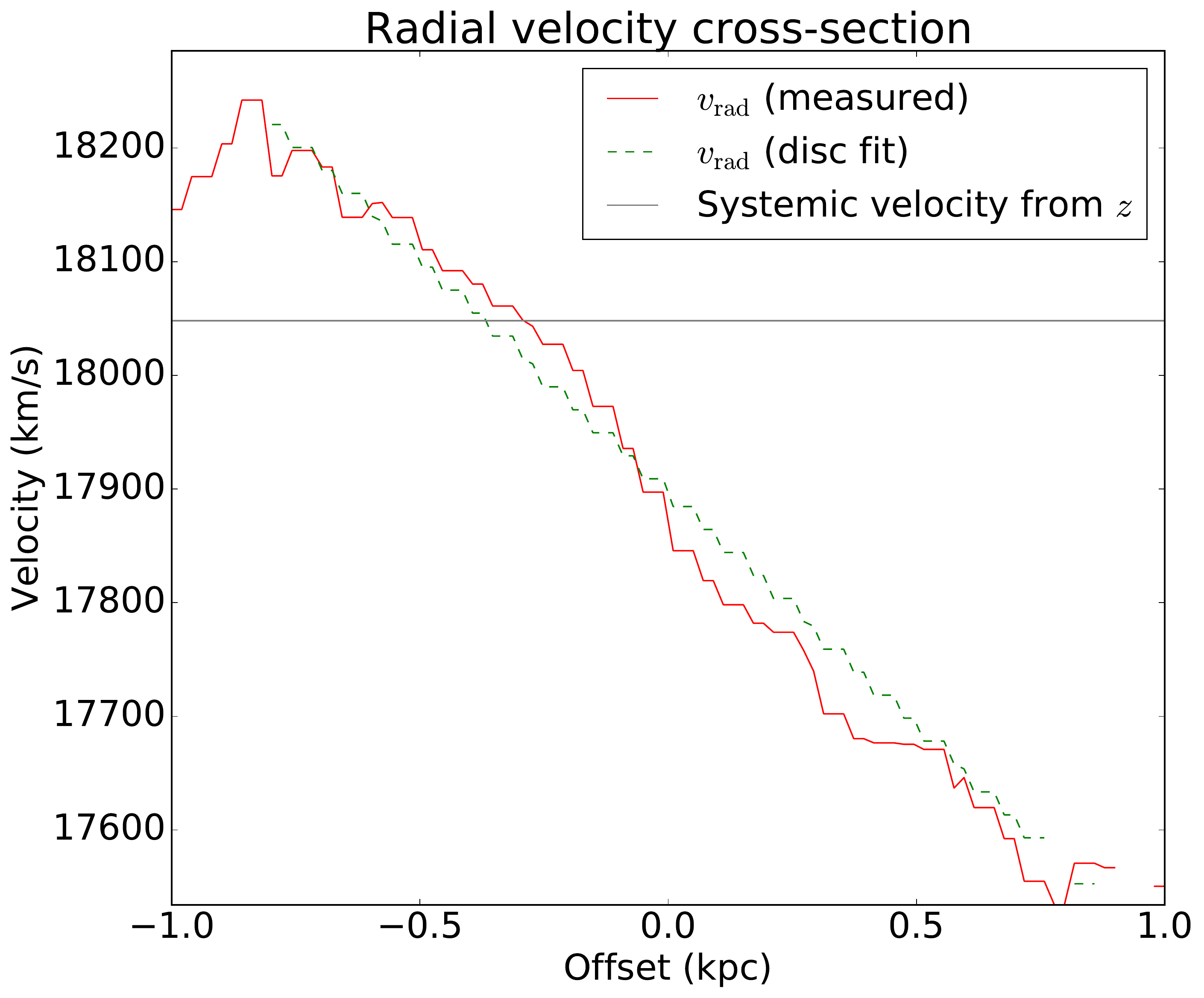}
	\caption{A cross section of the measured (red solid line) and model fit (green dashed line) radial velocity taken along the black dashed lines indicated in Fig.~\ref{fig: disc fit}. The systemic velocity of the galaxy, as estimated from the redshift, is indicated by the horizontal line, showing that the \hh{} emission has a systemic blueshift of $\approx 150\,\rm km\,s^{-1}$.}
	\label{fig: disc fit PV diagram}
\end{figure}

To determine whether the \hh{} belongs to the circumnuclear disc, we fit a simple disc model to our data using \textsc{mpfit}. Our model has solid-body rotation out to a break radius $r_b$, and a flat rotation curve for $r > r_b$, and we fit the systemic velocity as a free parameter. We do not account for beam smearing in the fit. Fig.~\ref{fig: disc fit} shows the model fit and residuals, and Fig.~\ref{fig: disc fit PV diagram} shows a cross-section of the radial velocities taken along the dashed black line shown in Fig.~\ref{fig: disc fit}.
The disc fit has a PA approximately perpendicular to the jet axis, indeed consistent with that of the circumnuclear disc detected in previous studies. 
The fitted inclination of the disc is consistent with the Easternmost edge being closest to us, which is also consistent with the greater \hi{} opacity of the Eastern radio lobe\,\citep{Conway1996,Struve2012}. At the edge of the disc ($r \approx 0.8\,\rm kpc$), the rotational velocity $v_c \approx 425\,\rm km\,s^{-1}$ with respect to the systemic velocity is comparable to that of the HCO$^+$ emission\,\citepalias{GarciaBurillo2007}. 
Therefore, our disc fit shows the warm \hh{} probes the interior $\sim \rm kpc$ of the circumnuclear disc of 4C\,31.04, where the apparent solid-body rotation is a result of the warm \hh{} not extending far enough into the disc to reach the turnover in the rotation curve.

Interestingly, the \hh{} emission has a net blueshifted velocity of $\approx 150 \rm\,km\,s^{-1}$ relative to the systemic velocity derived using the redshift ($z = 0.0602$), which is markedly similar to that of \hco{}\,\citepalias{GarciaBurillo2007} and \hi{}\,\citep{Struve2012} viewed in absorption over the nucleus.

A uniform distribution of dust in the disc could cause the flux-weighted mean velocity to have a significant blueshift; however this scenario would require an unrealistically large $A_{K}$. 
We instead speculate that the warm \hh{} traces clouds of gas being radially accelerated by jet plasma percolating through the disc, and that the net blueshift is because redshifted gas on the far side of the disc is obscured by extinction. 

We note that the $\sim100\,\rm km\,s^{-1}$ uncertainty in the redshift could be the cause this apparent blueshift. However, adopting a lower redshift to force the \hco{}, \hi{} and \hh{} components to coincide with the galaxy's systemic velocity would result in the \feii{} emission in the nucleus being significantly redshifted. This would be difficult to explain under our interpretation that the \feii{} emission arises from the jet accelerating material out of the nucleus. We discuss this further in Section\,\ref{subsec: Explaining the peculiar blueshift}.

The velocity residuals of the disc fit (Fig.~\ref{fig: disc fit}) reveal redshifted and blueshifted velocity residuals to the NE and SW of the nucleus respectively which are similar to the radial velocity of the \feii{} emission (Fig.~\ref{fig:[Fe II] vrad}), suggesting the kinematics of the molecular disc could be disrupted by the same processes causing the \feii{} emission. However, referring to the \textit{HST} $R-H$ image (Fig.~\ref{fig: HST overlays}) and to the flux distribution of the warm \hh{} (Fig.~\ref{fig:H2 flux}), the circumnuclear disc is clearly warped. The residuals in Fig.~\ref{fig: disc fit} could simply arise from our disc fit not accounting for this warp, and so this alone cannot be interpreted as evidence of kinematically disturbed molecular gas in the inner regions of the disc.

\subsubsection{Gas temperature}
\begin{figure}
	\centering
	\includegraphics[width=1\columnwidth]{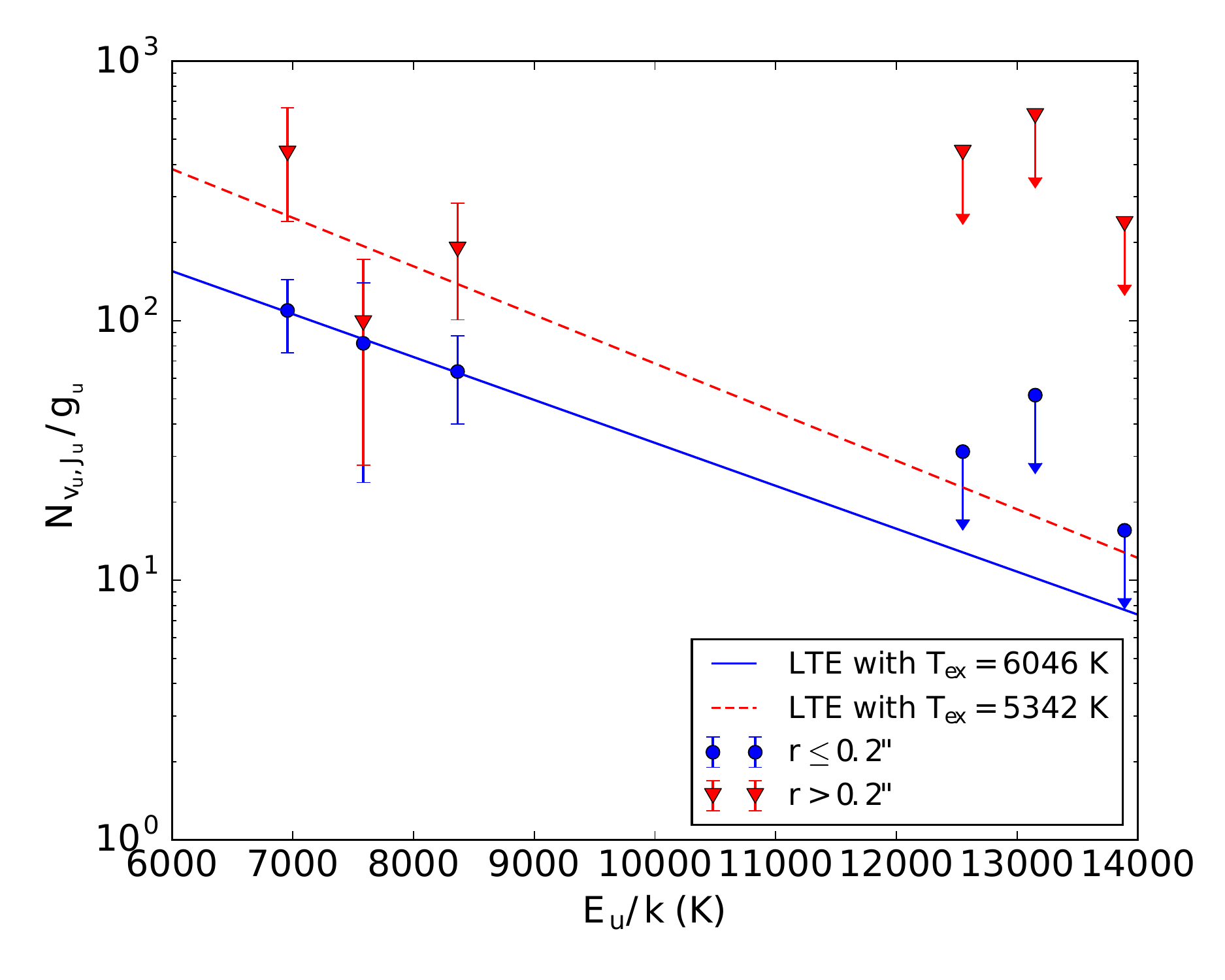}
	\caption{\hh{} excitation diagram of 4C\,31.04. Red triangles and blue circles indicate estimates of $N_\text{obs}(\nu_u,J_u) / g_u $ for the extended ($r > 0.2''$) and central ($r \leq 2''$  regions respectively. Arrows indicate emission lines for which we can only estimate upper limits.}
	\label{fig:excitation diagram}
\end{figure}

If \hh{} is in thermal equilibrium, the Boltzmann equation describes the level populations of the vibrational states. 
To determine whether or not the molecular gas is in thermal equilibrium, we use an excitation diagram, where we plot the level populations estimated from emission line fluxes as a function of the level energy (see Fig.~\ref{fig:excitation diagram}). This also enables us to constrain the temperature of the warm \hh{}. 
If the gas is in thermal equilibrium, the points will lie along a straight line in log space of level population versus transition energy with slope $-1/T_{\text{kin}}$ where $T_\text{kin}$ is the kinetic temperature of the gas. 
To convert emission line fluxes to level populations we use the method described in \citet{Rosenberg2013}: 

\begin{equation}
\frac{N_\text{obs}(\nu_u,J_u)}{g_u} = \frac{4\upi\lambda_{u,l}}{hc}\frac{I_\text{obs}(u,l)}{A(u,l)}\;,
\end{equation}
where $N_\text{obs}(\nu_u,J_u)$ is the observed column density of \hh{} molecules in the upper level $u$, $g_u$ is the statistical weight of the upper level, $\lambda_{u,l}$ is the rest-frame wavelength corresponding to the transition, $I_\text{obs}(u,l)$ is the measured flux of the transition and $A_{u,l}$ is the spontaneous emission coefficients, here obtained from \citet{Wolniewicz1998}. 

The excitation diagram for the \hh{}\,1--0\,S(1), S(2) and S(3) transitions in both the central and extended components of the \hh{} is shown in Fig.~\ref{fig:excitation diagram}, where we plot $N_\text{obs}(\nu_u,J_u)/g_u$ as a function of the temperature corresponding to the transition energy of each line. For non-detected emission lines we use the upper limits in these regions (Table~\ref{tab:measured line fluxes}). 
Unfortunately due to the close spacing in temperature of the \hh{}\,1--0\,S(1,2,3) transitions, and because we only have upper limits for the \hh{}\,2--1\,S(1,2,3) lines, we cannot place a very restrictive constraint on the temperature in the extended component of the \hh{}; however, the emission line fluxes in the central 0.4'' are consistent with a temperature $5000-6000\rm\,K$, indicating that the \hh{} in the central regions is much hotter than the $\sim 300\,\rm K$ \hh{} probed by \textit{Spitzer} observations\,\citep{Willett2010}.

%%%%%%%%%%%%%%%%%%%%%%%%%%%%%%%%%%%%%%%%%%%%%%%%%%%
\subsubsection{Excitation mechanism}
\label{subsubsec: H2 excitation mechanism}
Ro-vibrational \hh{} emission can trace collisionally excited gas processed by shocks that are not fast enough to dissociate \hh{} molecules.
These emission lines can also be emitted by \hh{} molecules excited by fluorescence from UV photons from O and B-type stars. 
We use two key emission line ratios (Table~\ref{tab: line ratios}) to demonstrate that shock excitation is the most likely scenario.

\begin{figure*}
	\centering
	\subcaptionbox{\label{fig: line ratio map H2}}{		
		\includegraphics[width=0.48\textwidth]{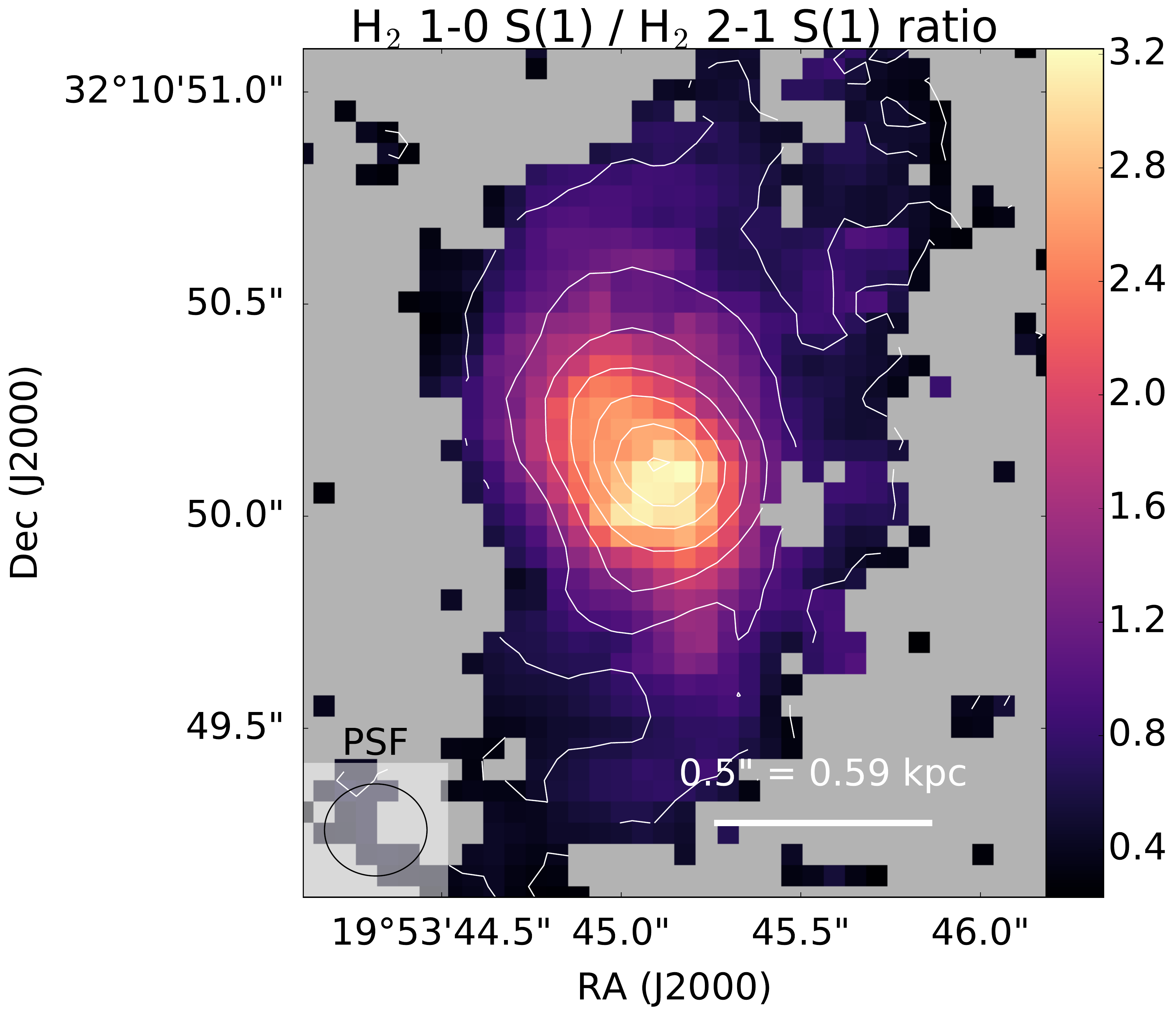}
	}
	\hfil
	\subcaptionbox{\label{fig: line ratio map Brgamma}}{		
		\includegraphics[width=0.48\textwidth]{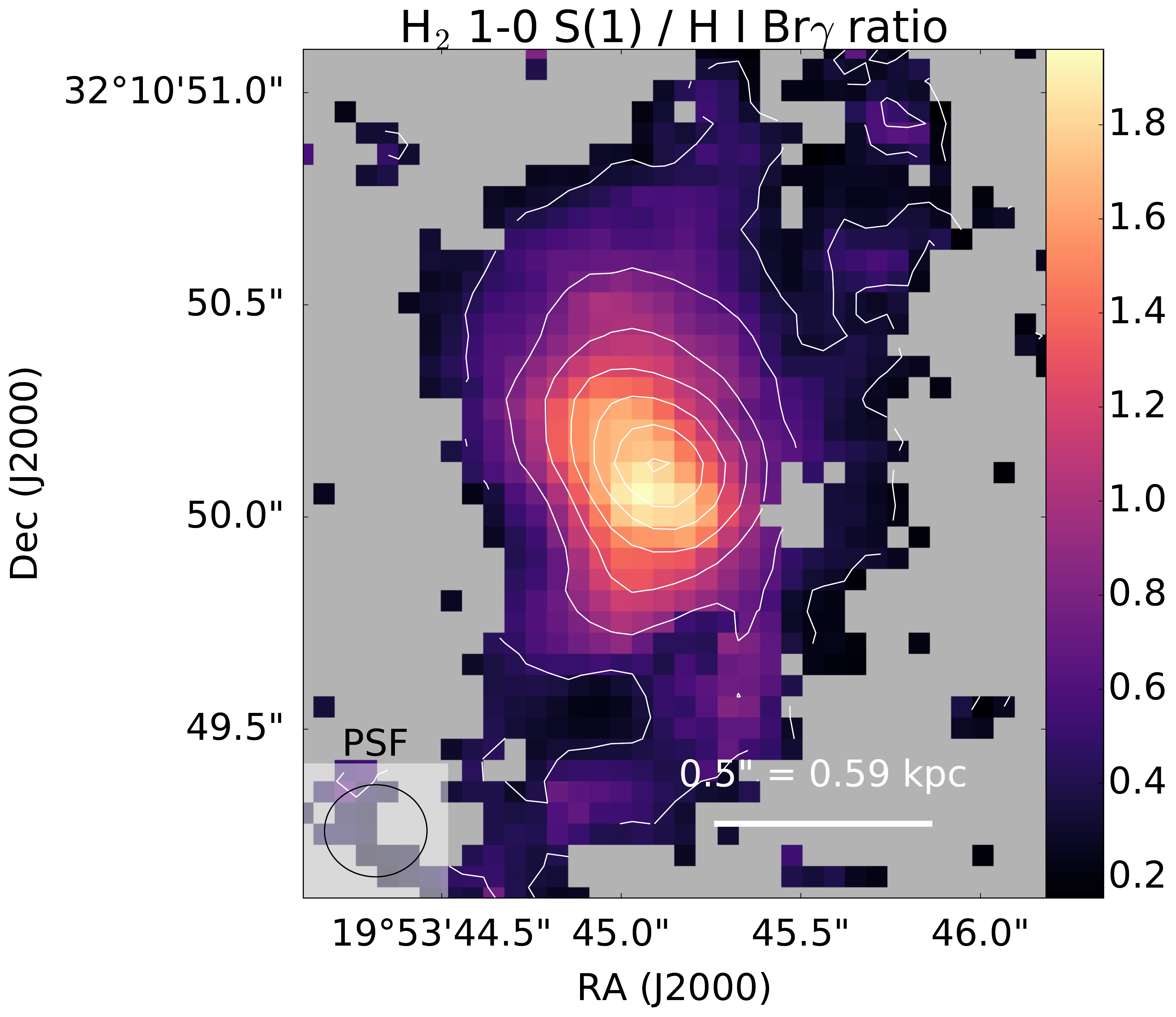}
	}
	\caption{Emission line ratio maps that we use to determine the excitation mechanism for the warm \hh{} in different regions. (a) shows the \hh{}\,1--0\,S(1)/\hh{}\,2--1\,S(1) ratio and (b) shows the \hh{}\,1--0\,S(1)/Br\,$\gamma$ ratio, and the white contours show the \hh{}\,1--0\,S(1) flux. The values of $\sim 3$ and $\sim 2$ in the inner 0.4'' in (a) and (b) respectively suggest shocks, and not star formation, are the likely excitation mechanism.}
	\label{fig: line ratio maps}
\end{figure*}

Shock-excited \hh{} can be distinguished from \hh{} excited by a UV stellar radiation field by the relative level populations: fluorescent excitation tends to populate higher-level $\nu$ states more than shock excitation. 
Therefore line ratios of ro-vibrational emission lines involving the same $J$ state transition with different upper $\nu$ levels can be used as indicators of the excitation mechanism.
In shock-heated gas, the ratio of the \hh{} 1--0/2--1\,S(1) lines tends to be much larger than in fluorescently-excited gas\,\cite[e.g.,][]{Nesvadba2011}. 
Fig.~\ref{fig: line ratio map H2} shows this ratio in each spaxel. We have used upper limits to estimate the 2--1\,S(1) flux; therefore these values should be interpreted as lower limits for the true line ratio. In the inner 0.4'', the ratio of the integrated fluxes of the two lines exceeds $\sim 2$, indicating shock excitation. 

At high enough densities ($n \gtrsim 10^3$\,cm$^{-3}$), where \hh{} is in LTE, level populations will be similar in both shock- and fluorescent-excited \hh{}, in which case the line ratios are misleading.
Here, we can eliminate fluorescent excitation by young stars because we do not detect Br\,$\gamma$, leaving shocks as the most likely mechanism.
Fig.~\ref{fig: line ratio map Brgamma} shows the ratio \hh{}\,1--0\,S(1)/Br\,$\gamma$ in each spaxel, where we have used upper limits to estimate the Br\,$\gamma$ flux. In the inner 0.4'', the ratio of the integrated fluxes of the two lines exceeds the 0.1--1.5 expected when the excitation source is UV heating in star-forming galaxies\,\citep{Puxley1990}. 
Additionally, our Br\,$\gamma$ upper limits are pessimistic, because we assume a Gaussian sigma determined from the combined H\,$\alpha$ + [N\,\textsc{ii}] equivalent width of \citet{Marcha1996} from an unresolved spectrum.
Therefore, combining this result with the high \hh{} 1--0/2--1\,S(1) ratio, we conclude that the ro-vibrational \hh{} emission is excited by shocks in the central region. 

In the extended component, we cannot rule out star formation as the source of excitation, although we note these results are not inconsistent with shock excitation, because both line ratios involve upper limits.

%%%%%%%%%%%%%%%%%%%%%%%%%%%%%%%%%%%%%%%%%%%%%%%%%%%
\subsubsection{Mass estimates}
We now estimate the dynamical mass and warm gas mass and compare our results with those from previous studies in Section\,\ref{subsec:The circumnuclear medium of 4C31.04}. 

Solid-body rotation in the \hh{} radial velocity implies the mass distribution interior to the disc can be approximated as a uniform-density sphere.
We estimate the enclosed dynamical mass using $M_\text{dyn} = v_c(r)^2 r/G$ where $v_c(r) \propto r$ is the rotational velocity at radius $r$ and $G$ is the gravitational constant. 
From our best-fitting thin-disc model, the rotational speed of the disc at its edge is $v_c(r = 0.8 \text{\,kpc}) \approx 425$\,km s$^{-1}$ and a dynamical mass $M_\text{dyn} \approx 3.4 \times 10^{10}\,\rm M_\odot$.

\citet{Dale2005} derived an expression for the mass of warm \hh{} using the \hh{}\,1--0\,S(1) flux $F_\text{1--0 S(1)}$ assuming LTE conditions and a temperature of 2000\,K:
\begin{equation}
M_{\text{\hh{},\rm\,warm}} \approxeq 5.08\,\rm M_\odot \left(\frac{F_\text{1--0 S(1)}}{10^{-16}\,\text{W m}^{-2}}\right) \left(\frac{d}{\text{Mpc}}\right)^2\;.
\label{eq:Warm H_2 mass from H_2 1--0 S(1) line flux (Dale et al. 2005)}
\end{equation}
Using this method we obtain $M_{\text{\hh{},\rm\,warm}}\,(T = 2000\,\rm K) = 9.7 \times 10^3\,\rm M_\odot$. 
We note that we are only able to place an upper limit of $\approx 6 \times 10^4\,\rm K$ on the \hh{} temperature using our excitation diagram (Fig.~\ref{fig:excitation diagram}), hence this mass is an approximate estimate.

\citet{Willett2010} also estimate the mass of warm \hh{} using mid-IR pure rotational (0--0) \hh{} emission lines from \textit{Spitzer} observations. For 4C\,31.04 they find $M_{\text{\hh{},\rm\,warm}} = (4.7 \pm 1.3) \times 10^{6}\,\rm M_\odot$ at a temperature $T = 338 \pm 100\rm\,K$, a mass $\sim 10^2$ greater than the mass of much warmer \hh{} we find. 
Combined with the temperature constraints provided by our excitation diagram (Fig.~\ref{fig:excitation diagram}), we conclude that the ro-vibrational emission we detect in our NIFS observations traces a small, and relatively hot, fraction of the total warm \hh{} in the nucleus of 4C\,31.04.

In Table~\ref{tab:Mass estimates} we compare the mass estimates of different ISM phases of 4C\,31.04 from this work and the literature.
We assume that the most reliable estimate of the cold \hh{} gas mass is that of \cite{OcanaFlaquer2010} derived from CO measurements, which gives a warm \hh{}-to-total molecular gas mass fraction of $\sim 10^{-3}$. 

\begin{table*}
	\centering
	\caption{Mass estimates of different gas phases in 4C\,31.04.
}
	\label{tab:Mass estimates}
	\begin{tabular}{c c c}
		\hline
		\textbf{Phase} & \textbf{Mass} $(\rm M_\odot)$ & \textbf{Reference}  \\
		\hline
		$M_{\rm BH}$ & $\leq 10^{8.16}$ & \citet{Willett2010} \\
		\hline
		$M_\text{dyn}\,(r < 0.8\,\rm kpc)$ & $3.4 \times 10^{10}$ & This work \\
		\hline
		$M_{\text{\hh{},\rm\,warm}}\,(\rm T = 2000\,\rm K)$ & $9.7 \times 10^3$ & This work \\
		$M_{\text{\hh{},\rm\,warm}}\,(\rm T = 338 \pm 100\rm\,K)$ & $(4.7 \pm 1.3) \times 10^{6}$ & \citet{Willett2010} \\
		$M_{\text{\hh{}}, \rm\,cold}$ & $(60.63 \pm 16.92) \times 10^8$ & \citet{OcanaFlaquer2010} \\
		\hline
		$M_\text{\hi{}}$ & $4.8 \times 10^7$ & \citet{Perlman2001} \\
		$N_\text{\hi{}}$ & $(1.2-2.4) \times 10^{21}\,\rm\,cm^{-2}$ & \citet{Struve2012}\\
		\hline
	\end{tabular}
\end{table*}

%It is clear that the galaxy has a very large reservoir of gas.
%, with an atomic to molecular gas mass ratio of $\sim 100$.
%\citetalias{GarciaBurillo2007} and \cite{OcanaFlaquer2010} use far-IR fluxes to estimate a dust mass which is then converted to a gas mass assuming a constant molecular gas-to-dust ratio of 100. 
%However, as we discuss in Section\,\ref{sec:Discussion}, the \feii{} emission shows that dust destruction is occurring in the nucleus of MCG\,5-4-18, meaning that the true molecular gas-to-dust ratio is probably higher than 100. 
%Additionally, neither \citetalias{GarciaBurillo2007} nor \cite{OcanaFlaquer2010} take into account AGN contamination in the far-IR fluxes, meaning both their estimates of $M_\text{dust}$ must be taken as upper limits.
%Therefore, we cannot confidently estimate the gas-to-dust ratio in MCG\,5-4-18. 

%%%%%%%%%%%%%%%%%%%%%%%%%%%%%%%%%%%%%%%%%%%%%%%%%%
\section{Discussion}\label{sec:Discussion}

In this section, we analyse the energetics, kinematics and morphology of the [Fe\,\textsc{ii}] and \hh{} emission and argue that they indicate strong jet-ISM interactions are occurring in 4C\,31.04.

Hydrodynamical simulations have shown that young, compact jets such as those in 4C\,31.04 are capable of influencing the evolution of their host galaxies by injecting turbulence and driving shocks into the ISM\,\citep[e.g.,][]{SutherlandBicknell2007,Wagner2011,Mukherjee2016,Mukherjee2018a,Mukherjee2018b}. Importantly, the coupling efficiency of the kinetic energy and momentum from the jet into the ISM peaks in these early stages of jet evolution\,\citep{Wagner2011,Mukherjee2016}, emphasising the importance of this epoch in the context of jet-driven feedback.

\begin{figure*}
	\centering
	\subcaptionbox
	{\label{fig: simulation midplane slice rho}}
	{\includegraphics[width=0.46\linewidth]{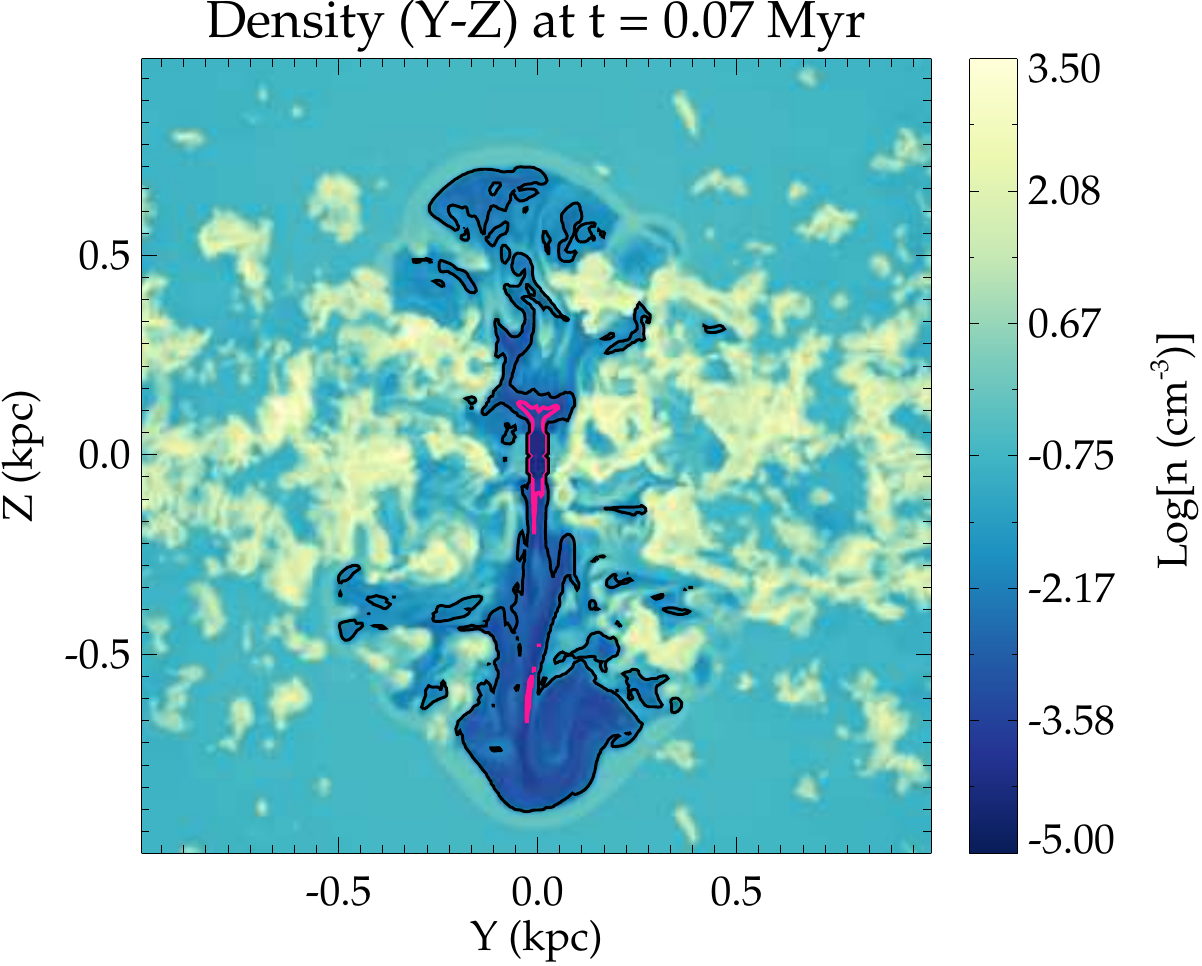}
	}
	\subcaptionbox
	{\label{fig: simulation midplane slice T}}
	{\includegraphics[width=0.46\linewidth]{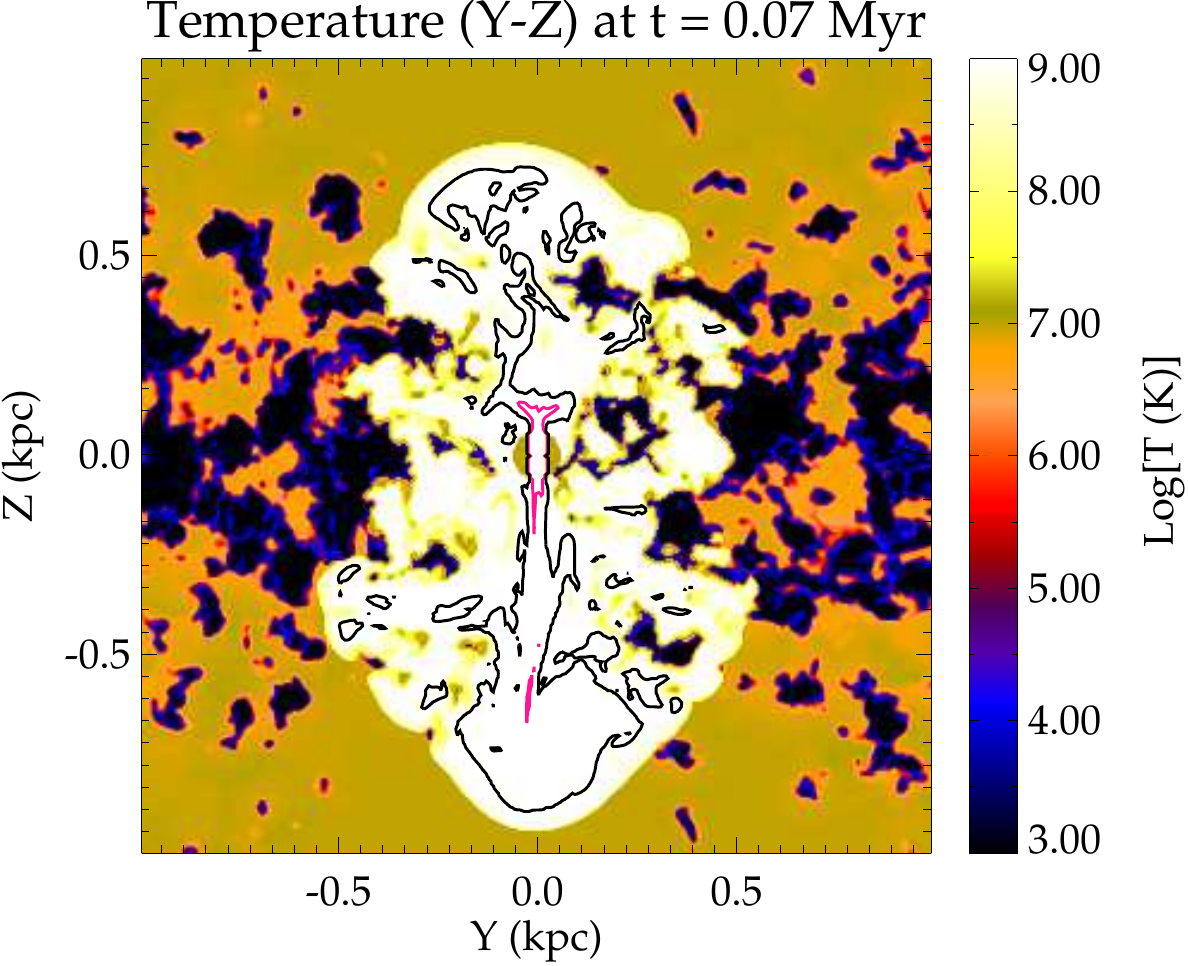}
	}
	\caption{Mid-plane slices from a hydrodynamical simulation of a jet with $F_{\rm jet} = 10^{45}\,\rm erg\,s^{-1}$ propagating in the $Z$-direction through a clumpy disc at an angle of $20^\circ$ (model C of \citet{Mukherjee2018b}; see their table 2 for simulation parameters); (a) shows the density and (b) shows the temperature. The magenta contours represent jet plasma that will emit brightly in the radio (jet tracer value $\phi = 0.5$) and the black contours trace much fainter plasma ($\phi = 0.005$) which fills the jet-driven bubble and drives shocks into the surrounding ISM.}
	\label{fig: simulation midplane slice}
\end{figure*}

\begin{figure*}
	\centering
	\subcaptionbox
	{\label{fig: synthetic radio image, DR=1e6}}
	{\includegraphics[width=0.48\textwidth]{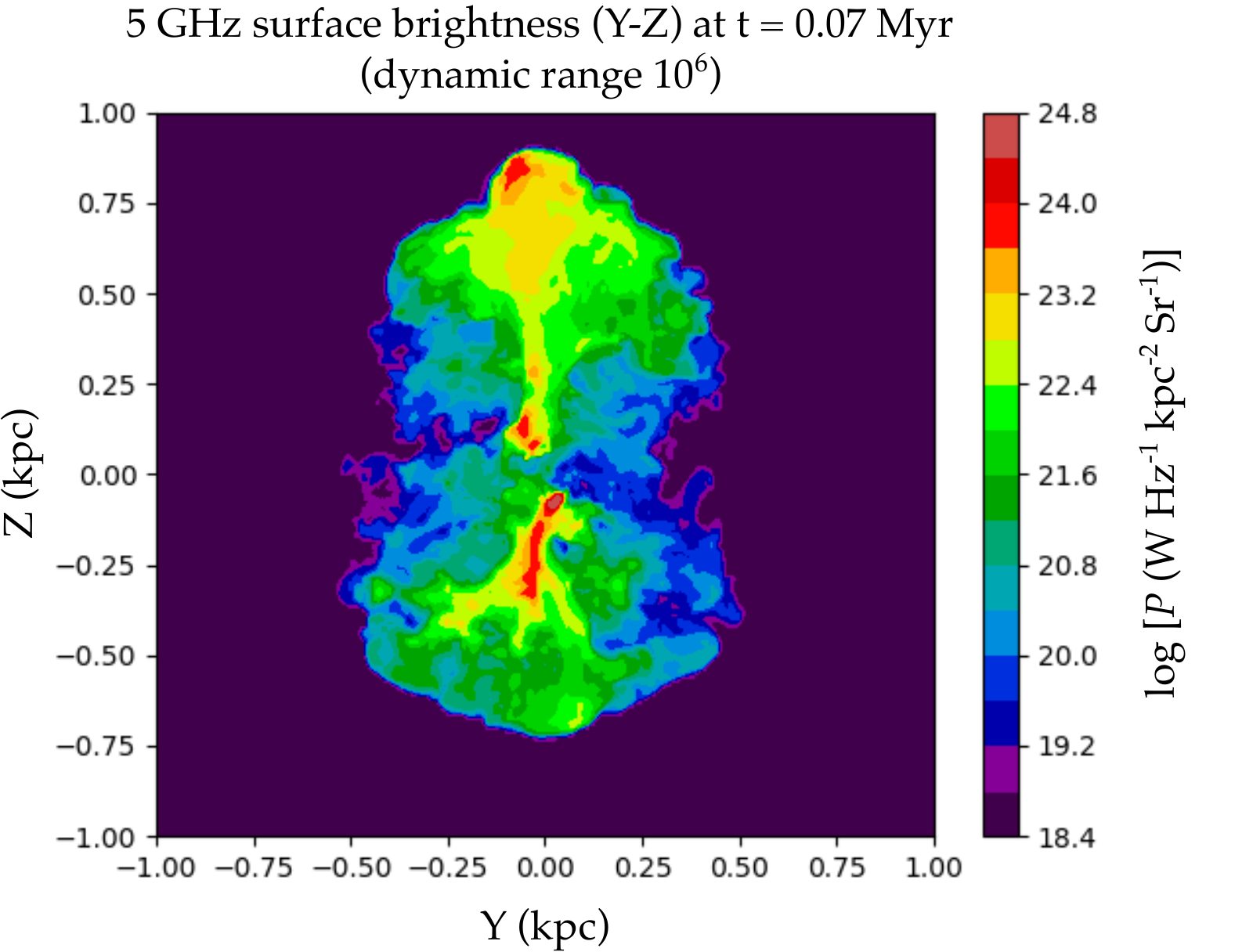}
	}
	\subcaptionbox{\label{fig: synthetic radio image, DR=1e2}}
	{\includegraphics[width=0.48\textwidth]{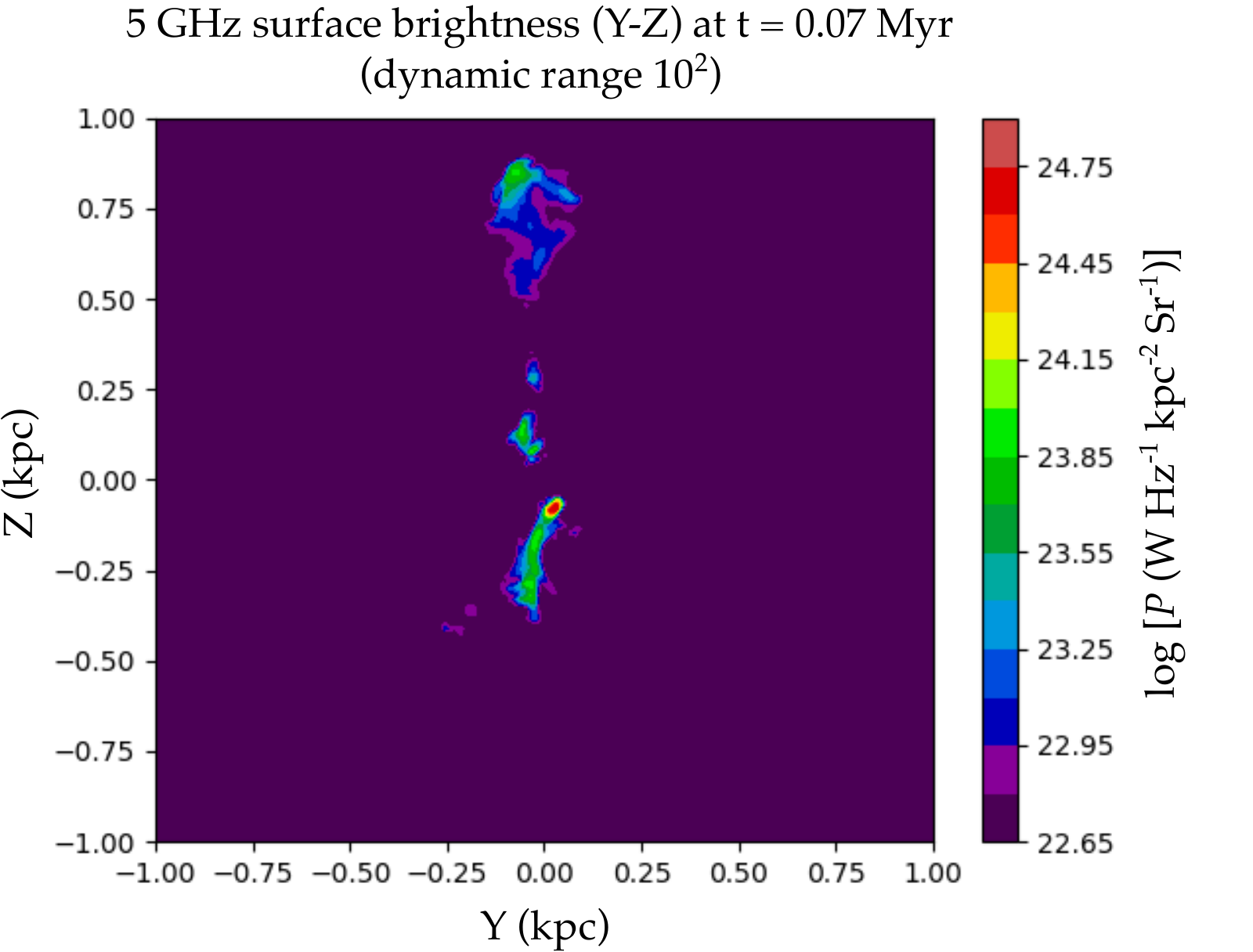}		
	}
	\caption{
		Synthetic radio surface brightness maps corresponding to the simulated mid-plane density and temperature slices shown in Fig.~\ref{fig: simulation midplane slice} at rest-frame 5\,GHz with dynamic ranges of $10^6$ and $10^2$ respectively. The axis labels labels are in kpc and the intensity is given in $\log\left[P (\rm W\,Hz^{-1}\,kpc^{-2}\,Sr^{-1})\right]$.
		Comparing (a) and (b) demonstrates that a high dynamic range may be necessary to observe low surface brightness plasma extending beyond the radio lobes into the jet-driven bubble. 
		We note that the dynamic range of (b) is comparable to that of the 5\,GHz VLBI observations of 4C\,31.04 by \citetalias{Giroletti2003}, suggesting that higher dynamic range observations could reveal jet plasma out to the extent of the \feii{} emission we observe in 4C\,31.04.}
	\label{fig: synthetic radio images}
\end{figure*}

In an early phase of jet evolution described as the `energy-driven bubble' stage by \citetalias{SutherlandBicknell2007}, the jets become deflected and split as they encounter dense clumps in the ISM, forming streams of plasma that percolate through the ISM over a broad solid angle.
Midplane density and temperature slices of a hydrodynamical simulation showing this phenomenon are shown in Fig.~\ref{fig: simulation midplane slice}, with corresponding synthetic radio images shown in Fig.~\ref{fig: synthetic radio images}. 
The jet plasma inflates a high pressure, pseudo-spherical, energy-driven bubble that drives a forward shock into the ISM, dispersing clouds and accelerating them outwards.
The low-density synchrotron-emitting plasma in the bubble cavity manifests as extended, low surface brightness radio emission\,\citep[e.g., fig. 5 in ][]{Wagner2011}. 
Simulations of jets propagating into dense, clumpy discs\,\citep{Mukherjee2018b} have shown that jets may also drive subrelativistic flows of plasma into the disc plane, inducing shocks and turbulence.
The jet plasma ablates clouds, and triggers hydrodynamical instabilities that form filaments.
Ram pressure and thermal pressure gradients from shocks accelerates these clouds and filaments in a radial direction, introducing significant non-circular motions into the disc.

\begin{figure*}
	\centering
	\includegraphics[width=0.75\linewidth]{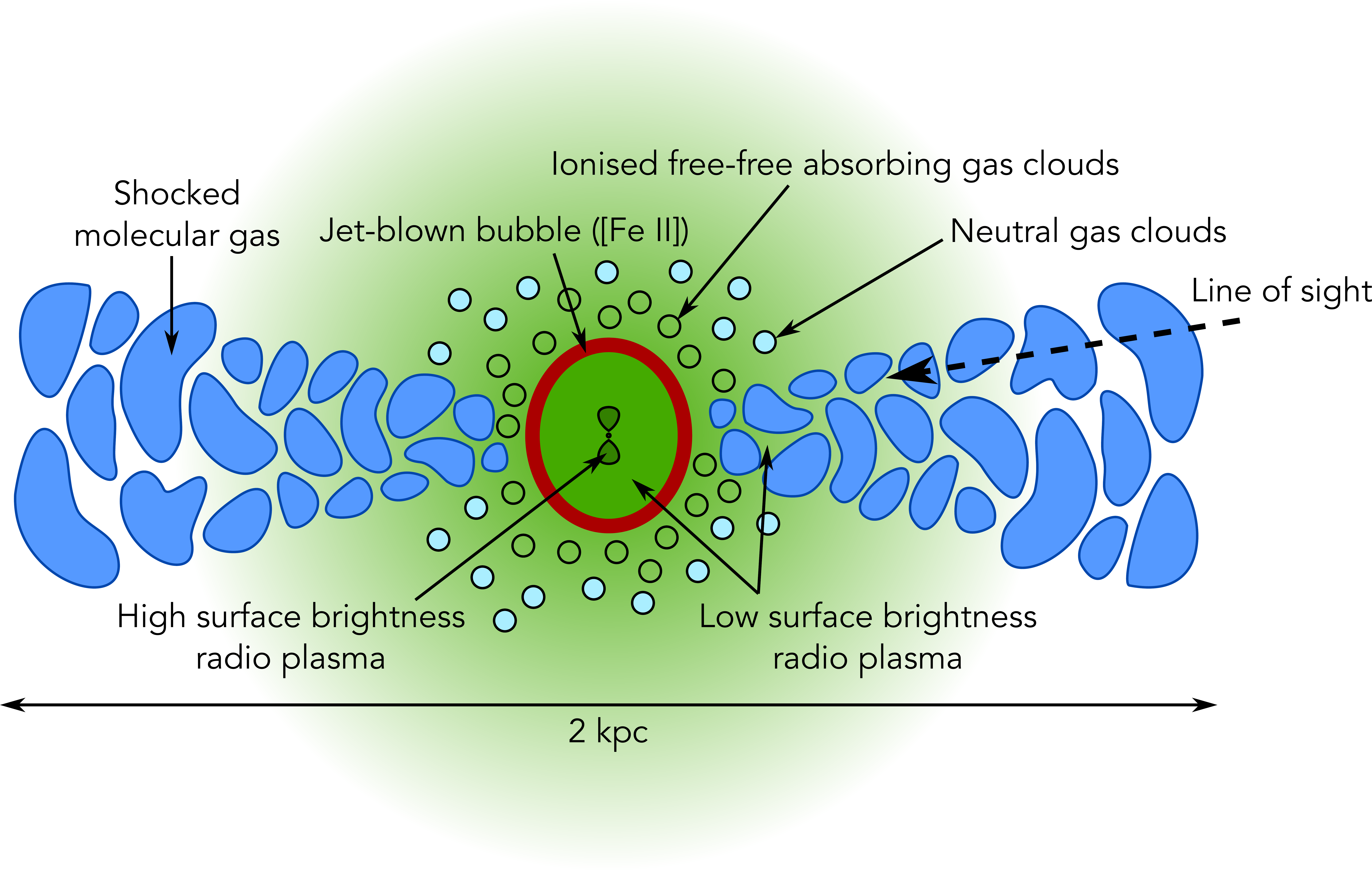}
	\caption{
		A top-down cross-section view of 4C\,31.04 showing different components of the shocked gas in context, approximately to scale. 
		Indicated in dark green are the 100\,pc-scale radio lobes visible in VLBI observations\,\citet{Giovannini2001,Giroletti2003}. 
		The paler green region represents low surface-brightness jet plasma filling the jet-driven bubble, which drives a shock into the surrounding gas and causes \feii{} emission (dark red). 
		Jet plasma also percolates radially through channels in the clumpy circumnuclear disc (blue), driving shocks into neutral gas and causing \hh{} emission.
		Clouds of ionized gas that free-free absorb synchrotron radiation from the jet plasma cause the spectral turnover at 400\,MHz are indicated in transparent circles. The pale blue circles represent clouds of neutral gas. 
		The line of sight is indicated by the dashed line; the disc is inclined such that the Western lobe is partially obscured by the disc, whereas the Eastern lobe is completely obscured by the disc, which is consistent with the \hi{} absorption map\,\citep{Conway1996,Struve2012}.}
	\label{fig:schematic}
\end{figure*}

% Observational signatures of flood-and-channel
Our observations, together with its young age and small size, strongly suggest that the radio jets of 4C\,31.04 are in the `energy-driven bubble' stage, where the jets are interacting strongly with the dense and clumpy circumnuclear disc. 
%the asymmetric and mottled appearance of the radio lobes, particularly the `hole' in the Eastern lobe, show that the jets are propagating into an inhomogeneous medium. 
%\hi{}, \hco{} and CO observations of its host galaxy indeed show that it has a large reservoir of gas concentrated in the circumnuclear regions. Moreover, 
By comparing our observations with hydrodynamical simulations, we formulate the model described by Fig.~\ref{fig:schematic}, which shows a top-down cross-sectional view of the circumnuclear disc of dust and atomic and molecular gas orbiting the nucleus. 
The inclination of the disc is such that the Eastern radio lobe is obscured; this is consistent with the greater \hi{} opacity on that side\,\citep{Conway1996,Struve2012}. 
The 100\,pc-scale radio lobes (dark green) are shown to scale. The jet plasma inflates an expanding bubble which drives fast shocks into the surrounding ISM, destroying dust grains and launching material out of the disc plane, which is traced by \feii{} (dark red) and high-velocity \hi{} clouds (pale blue circles) detected in absorption by \citet{Conway1996}. 
Shocked ionized gas (transparent circles) free-free absorbs synchrotron emission from the jet plasma, causing the spectral turnover at 400\,MHz. 
The jet plasma also drives radial flows into the disc, decelerating as it shocks molecular gas (blue), causing ro-vibrational \hh{} emission. The plasma may also radially accelerate this gas to speeds $\sim 100\,\rm km\,s^{-1}$, giving rise to non-circular motions including blueshifted \hh{} emission and \hco{} and \hi{} absorption.

\subsection{Updated jet flux estimate}\label{subsec: jet flux estimate}
Comparing the jet flux with observed emission line luminosities is important in determining whether it is plausible for the jets to be causing the line emission.
Hence, before we discuss the excitation mechanisms for the \feii{} and \hh{} emission in Sections\,\ref{sec: origin of FeII emission} and \ref{sec: origin of H2 emission} respectively, we provide an updated estimate of the jet flux of 4C\,31.04. 

% Jet flux calculations 
\citetalias{SutherlandBicknell2007} estimate the jet flux in 4C\,31.04 by calculating the minimum energy density in the radio lobes to produce the observed synchrotron flux in 1.7\,GHz VLBI observations, and find $4.4 \times 10^{43}$\,erg\,s$^{-1}$ and $1.5 \times 10^{43}$\,erg\,s$^{-1}$ in the Western and Eastern lobes respectively.
We follow the same minimum energy method to calculate the jet flux, this time using higher resolution VLBI observations of \citetalias{Giroletti2003} to achieve an improved estimate. Our input and output parameters are shown in Table~\ref{tab: parameters used in minimum energy calculation}.
Using their 5\,GHz VLBA image, we divide both East and West lobes up into sections of approximately constant flux density.
We model each section as volume with depth $L$ along the line of sight, which we assume to be equal to its width. We assume the volumes contain a randomly oriented magnetic field of strength $B$ and a relativistic electron population with distribution $N(\gamma) = K\gamma^{-a}$ for $\gamma \in [\gamma_1, \gamma_2]$. 
In this simple model, the minimum energy density required to produce a given specific intensity of synchrotron emission $I_\nu$ is 
\begin{equation}
\varepsilon_{\rm min,tot} = (1 + c_E) \epsilon_{e,\rm min} + \frac{B_{\rm min}^2}{2 \mu_0}
\label{eq: minimum energy total}
\end{equation}
where $\epsilon_{e,\rm min}$ is the minimum energy density in relativistic electrons and $c_E$ represents the energy fraction in other species, which we assume to be 0. 
The minimum energy magnetic field and energy density in particles are given by 
\begin{equation}
\begin{aligned}
B_{\rm min} &= \frac{m_e}{e}\left[ \frac{a + 1}{2} (1 + c_E) C_2(a)^{-1} \frac{c}{m_e} \left( \frac{I_\nu \nu^{\alpha}}{L} f(a, \gamma_1, \gamma_2) \right) \right]^{\frac{2}{a + 5}}\\
\varepsilon_{p,\rm min} &= \frac{4}{a + 1} \frac{B_{\rm min}^2}{2\mu_0}\\
\end{aligned}
\end{equation}
respectively, where $\alpha = (a - 1)/2$ is the spectral index of the synchrotron emission and $C_2(a)$ and $f(a, \gamma_1, \gamma_2)$ are constants. 
We find $\varepsilon_{p,\text{min}} \sim 10^{-5}\,\rm erg\,cm^{-3}$ and $B_{\text{min}} \sim 10^{-2}\,\rm G$ in all parts of the radio lobes.

% Calculating the jet flux
We assume that that half of the energy injected by the jets goes into $pdV$ work on the surrounding ISM and the other half into the energy density in both particles and magnetic field, which is typical of jet-driven bubble models\,\citep{Bicknell1997}. Under this assumption, the total energy in the lobes is given by
\begin{equation}
E_{\rm lobe} = \frac{1}{2} \sum_{{\rm min},i} \varepsilon_i V_i
\end{equation}
and the corresponding jet flux, assuming a constant rate of energy injection over a time $t_{\rm lobe}$, is given by
\begin{equation}
F_{\rm jet} = \frac{1}{t_{\rm lobe}} \frac{1}{2} \sum_{{\rm min},i} \varepsilon_i V_i
\end{equation}
where $V_i$ is the volume and $\varepsilon_i$ is the minimum energy computed using Eq.\,\ref{eq: minimum energy total}. 
Assuming an upper limit for the age of the lobes $t_{\rm lobe} = 5000\,\rm yr$ from \citetalias{Giroletti2003} based on synchrotron spectral decay, we find $F_{\rm jet,\,Western\,lobe} = 1.44 \times 10^{44}\,\rm erg\,s^{-1}$ and $F_{\rm jet,\,Eastern\,lobe} = 1.50 \times 10^{44}\,\rm erg\,s^{-1}$. 

\begin{table}
	\caption{Parameters used in determining the jet flux. Output parameters are denoted with daggers ($^\dagger$).}
	\begin{tabular}{c c c}
		\hline
		\textbf{Parameter} & \textbf{Symbol} & \textbf{Value} \\
		\hline
		Min. electron Lorentz factor & $\gamma_1$ & $10^2$ \\
		Max. electron Lorentz factor & $\gamma_2$ & $10^5$ \\
		Age of radio lobes\,\citepalias{Giroletti2003} & $t_{\rm lobe}$ & $5000\,\rm yr$ \\
		Temperature of ambient ISM & $T_a$ & $10^7\,\rm K$ \\
		Density of ambient ISM & $n_a$ & $0.1\,\rm cm^{-3}$ \\
		Radius of jet-blown bubble & $R_{\rm [Fe\,\textsc{ii}]}$ & $175\,\rm pc$\\ 
		\hline
		Eastern jet flux$^\dagger$ & $F_{\rm jet,\,Eastern\,lobe}$ & $1.50 \times 10^{44}\,\rm erg\,s^{-1}$\\
		Western jet flux$^\dagger$ & $F_{\rm jet,\,Western\,lobe}$ & $1.44 \times 10^{44}\,\rm erg\,s^{-1}$\\
		Total jet flux & $F_{\rm jet}$ & $2.94 \times 10^{44}\,\rm erg\,s^{-1}$ \\
		\hline
	\end{tabular}
	\label{tab: parameters used in minimum energy calculation}
\end{table}

%%%%%%%%%%%%%%%%%%%%%%%%%%%%%%%%%%%%%%%%%%%%%%%%%%%%%%%%%%%%%%%%%%%%%%%%%%%%%%%%%%%%%%%%%%%%%%%
\subsection{The origin of the \feii{} emission}\label{sec: origin of FeII emission}
Fast shocks destroy dust grains, releasing Fe\,\textsc{i} into the gas phase, which then becomes singly ionized by the interstellar radiation field. In the post-shock region, Fe\,\textsc{ii} becomes collisionally excited and emits emission lines in the near-IR, including \feii{}, which can therefore be used as a shock tracer. 

Fig.~\ref{fig:[Fe II] maps} shows that the \feii{} emission is localised to the innermost few 100\,pc of the nucleus, and that the radial velocity field is consistent with material being ejected from the disc plane on either side.
The kinematics and location of the \feii{} emission suggest that it arises from shocks driven by the expanding bubble inflated by the radio jets.
%The \feii{} emission could be a result of (1) supernovae (SNe) explosions, (2) radiation from the AGN, or (3) shocks induced by the radio jets.
%\todo{We can't say anything meaningful about whether AGN radiation could be causing it...}

% Supernovae
However, SNe explosions can also give rise to strong \feii{} emission; we now determine whether this is a plausible mechanism in 4C\,31.04.
To calculate the required SNe rate to produce the observed [Fe\,\textsc{ii}] emission, we use the empirical relationship between SN rate and [Fe\,\textsc{ii}]$_{1.26\,\rm \mu m}$ luminosity for starburst galaxies derived by \cite{Rosenberg2012}: 
\begin{equation}
\log \frac{\nu_{\text{SNrate}}}{\text{yr}^{-1}} = (1.01 \pm 0.2) \log \frac{L({\text{[Fe\,\textsc{ii}]}_{1.26\,\rm \mu m}})}{\text{erg s}^{-1}} - (41.17 \pm 0.9)
\label{eq: SN rate from [Fe II]}
\end{equation}
Because [Fe\,\textsc{ii}]$_{1.26\,\rm \mu m}$ lies outside the wavelength range of our observations, we assume any reddening is negligible and use the intrinsic ratio [Fe\,\textsc{ii}] 1.26/1.64\,$\mu$m $= 1.36$.
The integrated luminosity is given by $L({\text{[Fe\,\textsc{ii}]}_{1.644\,\rm \mu m}}) = 4 \upi D_L^2 F({\text{[Fe\,\textsc{ii}]}_{1.644\,\rm \mu m}}) = 1.19 \times 10^{40}\text{ erg s}^{-1}$ where $F({\text{[Fe\,\textsc{ii}]}_{1.644\,\rm \mu m}})$ is the integrated flux (Table~\ref{tab:measured line fluxes}) and the luminosity distance $D_L = 271.3$\,Mpc. 
Using Eq.\,\ref{eq: SN rate from [Fe II]} yields an integrated SN rate $\nu_\text{SN rate, [Fe\,\textsc{ii}]} = 0.2761 \rm\,yr^{-1}$. 

We now use the measured SFR to estimate $\nu_{\text{SNrate}}$ using a solar metallicity Starburst99\,\citep{Leitherer1999} model with a continuous $1\,\rm \rm M_\odot\,\rm yr^{-1}$ SF law and a Salpeter IMF at an age of $1\,\rm Gyr$. After multiplying to match the SFR of 4C\,31.04\,\citep[$4.9\,\rm M_\odot\rm\,yr^{-1}$,][]{OcanaFlaquer2010} we find $\nu_{\text{SNrate,  SFR}} = 0.1 \rm\,yr^{-1}$, less than half the rate required to power the \feii{} emission. 

%\note{This argument doesn't really stand up anymore because of the low line ratio...}
%We now rule out the AGN continuum radiation as causing the \feii{} emission. 
%To achieve our large observed ratio of \feii{}/Br\,$\gamma$, the AGN continuum must singly ionize Fe\,\textsc{i} whilst maintaining a low ionized fraction of hydrogen, which is only possible if the continuum is simultaneously weak and hard.
%Additionally, to destroy dust in order to release Fe into the gas phase without invoking shocks, the AGN must have a strong X-ray component. 
%Although there are currently no X-ray flux measurements of 4C\,31.04, we can rule out excitation by the AGN continuum on the grounds that a radiation field weak enough to reproduce our observed line ratio would be unable to produce the \feii{} luminosity without an unusually dense column.

We therefore rule out star formation as the excitation mechanism for the \feii{}, and instead argue that the emission is driven by a jet-ISM interaction. 
As illustrated in Fig.~\ref{fig:schematic}, the bubble drives fast shocks into the ISM, destroying molecular gas. The shocked gas is accelerated outwards by the forward shock and also by the jet streams, creating an expanding bubble illuminated in \feii{}.

%%%%%%%%%%%%%%%%%%%%%%%%%%%%%%%%%%%%%%%%%%%%%%%%%%%%%%%%%%%%%%%%%%%%%%%%%%%%%%%%%%
\subsection{The origin of the \hh{} emission}\label{sec: origin of H2 emission}
Large masses of warm \hh{} seem to be common in the hosts of nearby radio galaxies, suggesting a link between the \hh{} emission and radio activity\,\citep[e.g.,][]{Nesvadba2010,Ogle2010}. 
As discussed in Section\,\ref{sec:H2 emission}, we detect ro-vibrational \hh{} emission in the circumnuclear disc $\approx 2\,\rm kpc$ in diameter, which probes relatively hot ($\sim 10^3\,\rm K$) molecular gas. 
%As discussed in Section\,\ref{sec:H2 emission}, we find two morphological components of the \hh{} emission: an extended component $\approx 2\,\rm kpc$ in diameter, and a central component in the inner 0.4''. The ro-vibrational emission probes relatively hot ($\sim 10^3\,\rm K$) gas representing only a very small fraction of the total warm \hh{} mass. 
The line ratios show the warm \hh{} is shock excited (Table~\ref{tab: line ratios} and Fig.~\ref{fig: line ratio maps}). 

In this section, we show that the $\sim 10^3\rm\,K$ \hh{} represents a relatively hot fraction of a much larger reservoir of \hh{} heated by a jet-ISM interaction, the bulk of which is much cooler ( $\sim 10^2\,\rm K$). 
In our model, shown in Fig.~\ref{fig:schematic}, the jet-driven bubble drives fast shocks into the gas at small radii, causing \feii{} emission, before decelerating as it drives shocks into the denser molecular gas in the disc. 
We also postulate that jet plasma percolating radially throughout the disc also accelerates clouds to the observed systemic blueshift of $\approx 150\,\rm km\,s^{-1}$ in the warmer \hh{} component (Fig.~\ref{fig:H2 vrad}). 

\subsubsection{Excitation mechanism}
The ro-vibrational-emitting \hh{} probed by our observations is very warm (Fig.~\ref{fig:excitation diagram}), and cools rapidly. This phase is therefore very short-lived, and accordingly only represents a very small fraction of the total \hh{} mass (Table~\ref{tab:Mass estimates}). 
%Assuming that the ro-vibrational lines are powered by the dissipation of turbulent energy, the cooling time of this \hh{} is given by $\tau_{\rm diss} = \frac{3}{2} M_{\rm H_2} \sigma_{\rm H_2}^2 / L_{\rm H_2} \sim 10^3 \rm yr$. 
%However, in section x, we argue that if the extended \hh{} is heated by a jet-ISM interaction, the jets must have been active for \,100x longer. 
\citet{Willett2010} report a much larger ($\sim 10^6 \rm\,M_\odot$) reservoir of  cooler \hh{}, at a temperature $\sim 300 \rm\,K$.
We wish to determine whether this cooler component represents gas in the circumnuclear disc that has been processed by the jets and since cooled; to do this we use mid-IR diagnostics to reveal the excitation mechanism of the cooler \hh{} component.

% Demonstrating that the cooler H2 component is shock excited
Using the combined luminosity of the \hh{}\,0--0\,S(0,1,2,3) lines\,\citep{Willett2010} we place a lower limit on the ratio $L(\rm{H_2}) / L(\rm{PAH\,7.7\,\mu m}) \geq 0.1$, where we calculate the luminosity of the PAH feature at $7.7\,\mu \rm m$ using that of the $11.3\,\mu \rm m$ and assuming $L(\rm{PAH\,6.2\,\mu m})/L(\rm{PAH\,7.7\,\mu m}) = 0.26$, which holds for the sample of \hh{}-luminous radio galaxies of \citet{Ogle2010}; this gives $L(\rm PAH,\,7.7) = 1.55 \times 10^{42}\,\rm erg\,s^{-1}$. The criterion of \citet{Ogle2010} ($L(\rm{H_2}) / L(\rm{PAH\,7.7\,\mu m}) > 0.04$) strongly suggests the \hh{} is shock heated.

We also use the diagnostic diagram of \citet[][their fig. 6]{Nesvadba2010}, which separates star formation from other mechanisms as the source of \hh{} heating using the luminosity ratios of the summed \hh{}\,0--0 S(0)--S(3) lines to the $^{12}\rm CO(1-0)$ and to the PAH feature at $7.7\,\rm \mu m$. Whilst the CO emission traces cold molecular gas from which stars form, PAH features trace UV photons excited by star formation; hence these ratios can be used to indicate the contribution of star formation in photon-dominated regions (PDRs) to the \hh{} heating.
To calculate the CO(1-0) luminosity, we convert the $I_{\rm CO}$ given by \citet{OcanaFlaquer2010} into a luminosity using eqn. 3 of \citet{Solomon1997}, yielding $L(\rm CO) = 3.56 \times 10^{38}\,\rm erg\,s^{-1}$.
We find $L(\rm H_2) / L(\rm PAH,\,7.7) \geq 0.148$ and $L(\rm H_2) / L(\rm CO) \geq 6.43 \times 10^{2}$, placing 4C\,31.04 well outside the regions covered by PDR models, showing that the $\sim 300\,\rm K$ \hh{} component is not predominantly heated by UV photons, leaving shocks, cosmic rays and X-ray heating as plausible mechanisms. 
%We cannot rule out X-ray excitation because there are no X-ray observations of 4C\,31.04. 

% Arguing that it's probably shocks because of kinematic arguments: the weird blueshift
Based on our above arguments, both the $\sim 100\,\rm K$ and $\sim 10^3\,\rm K$ \hh{} is heated by a mechanism other than star formation. 
We argue that the $\sim 10^3\,\rm K$ and $\sim 100\,\rm K$ \hh{} are physically associated; in this case the strong shock signature we observe in the former (Fig.~\ref{fig: line ratio maps}) shows that the warm \hh{} is shock excited. 
%\todo{I think we need another sentence in here to argue why the two components should be physically associated.}

% Accretion?
\subsubsection{What is driving the shocks?}
\label{subsubsec: what is driving the shocks in the H2?}
We now show that the observed \hh{} luminosity cannot be produced by shocks driven by gas accreting on to the disc. 
The energy dissipated by gas accreting on to the disc at a rate $\dot{M}$ from $r = \infty$ to $r = r_0$ is given by 
\begin{equation}
L_{\rm disc} = \left[\frac{1}{2}v_c(r_0)^2 + \Phi(r_0)\right]\dot{M}
\label{eq: accretion disc luminosity}
\end{equation}
where $\Phi(r)$ is the galactic potential and $v_c(r)$ is the velocity of a circular orbit at radius $r$. We assume that the jet-driven bubble disrupts the \hh{} disc at smaller radii, hence we use $r_0 = 175\,\rm pc$, the approximate extent of the \feii{} emission.

To obtain $\Phi(r)$ and $v_c(r)$, we fit a simple $1/n$ Sersi\'c profile to the $K$-band continuum and use the analytical expressions given by \citet{Terzic&Graham2005}. 
We estimate the stellar mass $M_*$ from the 2MASS $Ks$-band magnitude using a simple stellar population model of \citet{Bruzual&Charlot2003} with a single instantaneous burst of star formation, assuming solar metallicity, a Chabrier IMF and an age of a few Gyr. This yields $\log M_* \approx 11.4$. 
To check the validity of our $M_*$ estimate, we compare the circular velocity predicted by the model with that of the warm \hh{} in the circumnuclear disc from our NIFS observations. We find that a lower stellar mass $\log M_* = 10.9$ is better able to reproduce the observed $v_c$ at $1\,\rm kpc$. However, we find that this correction only changes the resulting accretion disc luminosity to within an order of magnitude; hence, for simplicity, we use the stellar mass predicted by the 2MASS $Ks$-band magnitude.

If we assume a steady state accretion model, that is $\dot M$ is equal to the accretion rate onto the black hole, and that $10\rm\,per\,cent$ of the rest-mass energy of the accreted material per unit time is emitted in the form of the radio jets, i.e., 
$L_{\rm BH} = 0.1 \dot{M} c^2 = 2.94 \times 10^{44}\,\rm erg\,s^{-1}$,
then the accretion rate $\dot{M} \approx 0.05\,\rm M_\odot\,\rm yr^{-1}$. Using Eqn. \ref{eq: accretion disc luminosity}, we estimate an accretion disc luminosity $L_{\rm disc} \sim 10^{40}\,\rm erg\,s^{-1}$, an order of magnitude lower than the observed \hh{} 0--0 luminosity $L_{\text{\hh{}}} \geq 2.3\times10^{41} \rm \,erg\,s^{-1}$, enabling us to rule out accretion as the mechanism heating the \hh{}. Meanwhile, the \hh{} luminosity represents $\sim 0.1\rm\,per\,cent$ of the estimated jet flux (Section\,\ref{subsec: jet flux estimate}), so that there is ample jet power to drive the \hh{} luminosity via radiative shocks.
We therefore conclude that the shocks are being driven by the jets, a scenario which also explains the sharp peak in the \hh{} flux in the vicinity of the nucleus (Fig.~\ref{fig:H2 flux}).

\subsubsection{Explaining the peculiar blueshift in the \hh{}}\label{subsec: Explaining the peculiar blueshift}
Shocks induced by a jet-ISM interaction out to kpc-scale radii in the circumnuclear disc could explain the peculiar systemic blueshift of $\approx 150 \rm\,km\,s^{-1}$ we observe in the \hh{} (Fig.~\ref{fig:H2 vrad}) relative to the redshift $z = 0.0602$ of \citetalias{GarciaBurillo2007}.

Previous observations have revealed cold gas components with remarkably similar blueshifts: \citetalias{GarciaBurillo2007} detected an unresolved \hco{} component in absorption over the nucleus with a systemic blueshift of $\approx 150 \rm\,km\,s^{-1}$, whilst \citet{Struve2012} found that the centre of the integrated \hi{} absorption profile is centered on a similar blueshifted systemic velocity. 
The absorption profile also has a blue wing (their fig. 7) extending to $\sim 200\,\rm km\,s^{-1}$. Prominent blue wings in the \hi{} absorption profiles of powerful radio galaxies are a signature of fast neutral outflows\,\citep{Morganti2005}; we speculate that the blue wing in the profile of 4C\,31.04 indicates that the jets are accelerating neutral gas out of the nucleus, albeit to much lower velocities than are observed in these more extended radio galaxies.

% Arguing that the redshift is correct
It is possible that the redshift is in fact lower than the estimate of \citetalias{GarciaBurillo2007}, and that the \hh{} rotation curve is centred on the systemic velocity of the galaxy.
%It is possible that the systemic velocity we measure for the \hh{} represents that of the host galaxy, and that the redshift is in fact lower than currently thought.
\citetalias{GarciaBurillo2007} calculated the redshift of 4C\,31.04 by assuming that the \hco{} emission components to the North and South of the nucleus represented gas rotating in the disc, and that the mean velocity of the two components corresponds to the galaxy's systemic velocity. \citet{Struve2012} argue that this assumption may be flawed if gas in the disc is dynamically unsettled. 
Correcting the redshift to remove the blueshift in the $\sim 10^3 \rm\,K$ \hh{}, \hi{} and \hco{} absorption yields $z = 0.0597 \pm 0.001$. Whilst this is consistent with the redshift from optical spectroscopy ($z = 0.060 \pm 0.001$, \citet{Marcha1996}), it would mean the \feii{} emission is in fact redshifted by $\approx 150 \rm\,km\,s^{-1}$, which would be difficult to explain under our interpretation that it traces a jet-driven bubble.

An alternate explanation is that the blueshifted warm and cold molecular phases trace clouds of gas being radially accelerated in circumnuclear disc by jet plasma. 
As mentioned earlier, hydrodynamical simulations by \citet{Mukherjee2018b} of jets evolving in galaxies with clumpy gas discs have shown that the expanding bubble driven by the jets drives subrelativistic radial flows into the disc plane.
These flows drive turbulence and shocks into the gas, and introduce significant non-circular motions. In addition to the observed blueshifted gas, radial flows of jet plasma may be able to explain the reported unrelaxed dynamics in the gas disc\,\citepalias{Perlman2001,GarciaBurillo2007}, which has previously been attributed to gas settling on to the disc in the process of accretion.
We note that $\sim 100\,\rm pc$-scale equatorial outflows have been recorded in other radio galaxies, such as NGC\,5929\,\citep{Riffel2015} and NGC\,1386\,\citep{Lena2015}. These outflows have tentatively been attributed to torus outflows or accretion disc winds; it is unclear whether this mechanism could drive the relatively low-velocity blueshifted gas we observe in 4C\,31.04. 

% Outflow: jet age based on outflow speed
We now calculate the kinetic energy associated with the observed blueshift in the \hh{} to show that it is plausible that it is driven by the jets.
Assuming that both the $\sim 100\rm\,K$ and $\sim 10^3\rm\,K$ \hh{} are accelerated to the observed blueshifted velocity, the cooler component\,\citep[$4.7 \pm 1.3 \times 10^{6}\,\rm \rm M_\odot$,][]{Willett2010} will dominate the kinetic energy. If the blueshifted clouds have a velocity $150\,\rm km\,s^{-1}$, then the associated kinetic energy is $\approx 1.1 \times 10^{54}\,\rm erg$. 
Assuming the gas is being pushed radially outwards at a constant velocity, the time taken for material to reach the farthest extent of the warm \hh{} disc $\approx 1\,\rm kpc$ from the nucleus is $\tau = 6.5 \times 10^6\,\rm yr$.
Assuming the jet has been accelerating this gas over this time period, this yields an energy injection rate $\sim 10^{41}\,\rm erg\,s^{-1}$, approximately $0.1\rm\,per\,cent$ of the jet power we estimated in Section\,\ref{subsec: jet flux estimate}, showing that it is indeed plausible. 

In light of these arguments, we tentatively agree with the redshift quoted by \citetalias{GarciaBurillo2007}, and speculate that the blueshifted \hh{}, \hi{} and \hco{} traces gas clouds being radially accelerated in the disc plane by the jet plasma. 
Confirmation of this scenario will require further observations, e.g., high-resolution optical spectroscopy to measure the galaxy's redshift using stellar absorption features.

% Inhibiting star formation
%\paragraph{Are the jets inhibiting star formation?}
%\todo{Ask Nicole if I'm calculating surface densities correctly.}
%Whilst the jet plasma can only accelerate clouds to a few 100 km\,s$^{-1}$ at kpc radii in the disc, and is therefore not able to eject gas from the galaxy's potential entirely, it may interrupt star formation by heating and driving turbulence into molecular gas. 
%To see if the jets are inhibiting star formation, we plot 4C 31.04 on the mid-IR Schmidtt-Kennicutt law of Nesvadba+2010, which plots the surface density of PAH emission against the surface density of cold molecular gas, representing available fuel for star formation. 
%We adopt Ocana-Falquer's estimate of the cold H2 mass from CO(1-0) observations (what is the beam size?) and find a surface density xyz. 
%For the PAH emission, we use the PAH luminosities given by Willett+2010, assuming a ratio xyz of 11.7 to 7.7, and for a lower limit of the surface density assume the emission is uniform over the area resolved by their observations (also need to get the slit size) . Using these two diagnostics, we find 4C 31.04 lies far below (how much?) the trend for star-forming galaxies, indicating that star formation is inhibited. Together with our shock diagnostics for the warm H2, we conclude that the jets are heating molecular gas in the disc and suppressing star formation.

% Inclined jet theory
\subsubsection{How far does the jet plasma extend?}
In the previous sections we have shown that both the \feii{} and \hh{} emission are caused by a jet-ISM interaction. 
However, the radio lobes extend $\approx 60\rm\,pc$ from the nucleus, whereas we detect \feii{} over a region $\approx 3$ times larger, and warm \hh{} out to $\sim \rm kpc$ radii--how could this emission possibly arise from a jet-ISM interaction?

%Percolating jet streams can extend beyond the jet lobes during the flood-and-channel phase.
Hydrodynamical simulations of jets propagating into clumpy discs \,\citep{Mukherjee2018b} show that the brightest regions of jet plasma may become temporarily frustrated by dense clouds in the disc, slowing its propagation. 
This effect becomes more pronounced when the jets are inclined with respect to the disc normal, as it increases the effective path length over which the jets interact with the dense ISM.
Meanwhile, the expanding bubble can advance rapidly once it escapes the dense ISM in the disc plane, allowing the bubble radius to grow several times larger than the radio lobes. 

Our observations show the jets in 4C\,31.04 are likely to be inclined $\sim 10^\circ - 20^\circ$ to the normal of the circumnuclear disc.
This geometry is supported by the inclination of $\sim 60^\circ$ we measure from the warm  \hh{} and by the jets being at an angle of $\lesssim 15^\circ$ with respect to the sky, with the Western lobe nearest\,\citep{Giovannini2001}. Moreover, the kinematics of the \feii{} emission line (Fig.~\ref{fig:[Fe II] vrad}) shows material being accelerated off the disc plane such that the Western lobe is pointing towards us.

%Hydrodynamical simulations of jets propagating into clumpy discs at a range of different angles\,\citep{Mukherjee2018b} show that varying the angle between the jet axis and the disc normal can have a dramatic effect on the outcomes of jet-ISM interaction.
%When the axis of the jets is inclined with respect to the disc normal, the jet tip is more likely to become temporarily frustrated by dense clouds in the disc, slowing its propagation. Meanwhile, the expanding bubble can advance rapidly once it escapes the dense ISM in the disc plane, allowing the bubble radius to grow several times larger than the radio lobes. 

%Our observations show the jets in 4C\,31.04 are likely to be inclined $\sim 10^\circ - 20^\circ$ to the normal of the circumnuclear disc.
%This geometry is supported by the inclination of $\sim 60^\circ$ we measure from the warm  \hh{} and by the jets being at an angle of $\lesssim 15^\circ$ with respect to the sky, with the Western lobe nearest\,\citep{Giovannini2001}. 
%Moreover, the kinematics of the \feii{} emission line (\ref{fig:[Fe II] maps}) shows material being accelerated off the disc plane such that the Western lobe is pointing towards us.

In order to illustrate the role that a dense and clumpy disc can play in determining the outcome of a jet-ISM interaction, we show midplane density and temperature slices from a hydrodynamical simulation in which the jets are inclined $20^\circ$ to the disc normal in Fig.~\ref{fig: simulation midplane slice}. In Fig.~\ref{fig: simulation midplane slice rho}, the brightest parts of the jet plasma (magenta contours), particularly in the +ve $Z$-direction, have become halted a short distance from the nucleus. Meanwhile, lower surface brightness plasma (black contours) propagates along channels in the clumpy ISM and fills the much larger bubble, which crucially may go undetected in high-resolution VLBI observations despite interacting strongly with the surrounding ISM. 
In Figs.~\ref{fig: synthetic radio image, DR=1e6} and \,\ref{fig: synthetic radio image, DR=1e2} we show corresponding synthetic 1\,GHz surface brightness maps with high and low dynamic ranges respectively, where we define the dynamic range as the maximum measured flux divided by the minimum flux level that can be detected. Comparing the two illustrates the importance of a high dynamic range in revealing the low surface brightness plasma that traces the true extent of the jet-driven bubble. 
We note that Fig.~\ref{fig: synthetic radio image, DR=1e2} is missing only $\approx1\rm\,per\,cent$ of the total flux recovered in Fig.~\ref{fig: synthetic radio image, DR=1e6}. This illustrates that VLBI observations with flux completeness measurements in excess of $99\rm\,per\,cent$ can miss jet plasma that may still be interacting strongly with the surrounding ISM.

Multiple flux completeness measurements of VLBI observations of 4C\,31.04 indeed indicates that some large-scale structure exists at lower surface brightnesses than have been observed. 
\citet{Cotton1995} find that 98\rm\,per\,cent and 76\rm\,per\,cent of the flux density of 4C\,31.04 measured with the VLA is recovered in VLBI observations at 1.7 and 8.4\,GHz respectively. \citetalias{Giroletti2003} find that approximately 90\rm\,per\,cent of the flux measured with single-dish observations is recovered with VLBI at 5\,GHz.
\cite{Altschuler1995} recover 80\rm\,per\,cent of the total flux density at 92\,cm. 
We note that the 5\,GHz VLBI image of \citetalias{Giroletti2003} has a dynamic range of $\sim 100$; this, combined with the $\sim 75 - 95\rm\,per\,cent$ flux completeness of GHz-range VLBI observations, suggests that 4C\,31.04 may indeed harbour low surface brightness radio emission out to the radii at which we observe shocked gas, resolving the inconsistency between the extent of the shocked gas and the radio lobes. 

%In this section we use radio observations to estimate both the age of the jet-blown bubble, by first estimating the jet flux, and to constrain the parameters of the density distribution in the ISM. 

% Age of the Fe II bubble
\subsection{Age of the radio source}
We cannot estimate the true age of the radio jets in 4C\,31.04 with existing VLBI observations as they resolve out low surface brightness radio emission that fills a much larger bubble revealed by our NIFS observations. 
Moreover, because the rate at which synchrotron-emitting electrons lose energy $E$ is proportional to $E^2$, high-frequency, high surface-brightness radio emission only probes the youngest synchrotron electrons; parts, or even most, of the emission from the jet plasma may be missed by GHz-frequency observations with a small dynamic range. 
We instead use the jet flux and our NIFS observations to estimate the true age of the radio jets. 

We model the radio lobes as bubbles expanding adiabatically into a uniform ISM using the model of \citet{Bicknell&Begelman1996}. We assume that the bubbles are expanding out of the disc plane into the ambient hot ISM with $p/k \sim 10^6\,\rm K\,cm^{-3}$ typical in the interiors of local elliptical galaxies\,\citep{Werner2012} and $T_a \sim 10^7\,\rm K$, $n_a = p / kT_a \approx 0.1\,\rm cm^{-3}$.
The age of the bubble is given by
\begin{equation}
t_b = \left( \frac{ 384 \upi }{ 125 } \right)^{1/3} \rho_a^{1/3} F_{\rm jet}^{-1/3} R_b^{5/3}
\end{equation}
where $R_{b}$ is the radius of the bubble and $F_{\rm jet}$ is the total jet flux in $\rm erg\,s^{-1}$.

We calculate a lower limit for the age of the radio source by assuming that the jet plasma has only reached the extent of the \feii{} emission, ignoring the extended \hh{} emission. In this case, we set $R_b \approx 175\,\rm pc$ and find $t_b \approx 17\,\rm kyr$, more than 3 times older than the age estimated using synchrotron spectral decay\,\citepalias[4000-5000\,yr,][]{Giroletti2003}. 

In Section\,\ref{subsubsec: what is driving the shocks in the H2?}, we estimated an upper limit for the jet age $6.5 \times 10^6\,\rm yr$ by calculating the time taken for material to reach the farthest extent of the warm \hh{} disc, and is an upper limit as the gas may have decelerated along its trajectory.

Although these age estimates are very crude, together they suggest that the previous age estimates based on VLBI imaging alone may not represent the true age of the source. Our results demonstrate the importance of optical or near-IR tracers of jet-ISM interaction in estimating the true extent of the jet plasma, particularly when existing radio observations have a low dynamic range or are not sensitive to the angular scales associated with the diffuse plasma filling the jet-driven bubbles. 

%Assuming that the jet-blown bubble is driving the \hh{} emission, we set $R_b = 1\,\rm kpc$ which yields $321.9\,\rm kyr$.
%If the jet plasma is accelerating molecular gas to the observed speeds, then we can also estimate jet age by calculating the time taken for material to reach the farthest extent of the warm \hh{} disc, $\approx 1\,\rm kpc$ from the nucleus, assuming it has a velocity of $\approx 150\,\rm km\,s^{-1}$. This gives $\tau = 6.5 \times 10^6\,\rm yr$, and is an crude upper limit as the gas may have decelerated along its trajectory. These two age estimates are broadly consistent with one another, showing that the kinematics of the equatorial outflow are consistent with our calculated age of the radio source.
%We also note that the cooling time of the $\sim 10^3\,\rm K$ \hh{} component is $\sim 10^3\,\rm yr$, suggesting that the jets are continually heating this gas.

\subsection{Density distribution of the clumpy ISM}
%Lastly, we use radio observations to constrain the parameters of the density distribution in the ISM, in order to inform future hydrodynamical simulations of 4C\,31.04. 
Free-free absorption (FFA) by clumpy gas ionized by the radio jets is a feasible cause of the spectral turnover in GPS and CSS sources\,\citep{Bicknell2018}. 
Here, we use the peak in the radio spectrum of 4C\,31.04 to infer the parameters of the density distribution of the ionized, free-free absorbing ISM, in order to inform future hydrodynamical simulations.
We use a simple analytical model to calculate the specific intensity of synchrotron-emitting plasma embedded in a clumpy free-free absorbing medium.

We assume the density $n$ of the ionized medium follows a log normal distribution, which is appropriate for a turbulent medium\,\citep[e.g.,][]{Federrath2012}.
The log normal distribution has the probability distribution function (PDF)
\begin{equation}
P(n) = \frac{1}{\sqrt{2 \upi}s}\exp{\left[-\frac{\left( \ln n - m\right)^2}{2 s^2}\right]}
\end{equation}
which is a Gaussian in $\ln n$, where $m$ is the mean log density and $s$ is the width of the distribution in log density.\,\citepalias[e.g.,][]{SutherlandBicknell2007}.
The parameters $s$ and $m$ can be related to the expected value $E(n)$ of the density, i.e. the mean density $\mu$, $E(n^2)$, and the variance $\sigma^2$ of the density distribution using the relations
\begin{equation}
\mu = E(n) = e^{ m + \frac{1}{2}s^2 }
\label{eq: E(n) of log-normal distribution}
\end{equation}
\begin{equation}
E(n^2) = e^{2(m + s^2)}
\label{eq: E(n2) of log-normal distribution}
\end{equation}
\begin{equation}
\sigma^2 = \mu^2 \left( e^{s^2} - 1\right).
\label{eq: sigma of log-normal distribution}
\end{equation}

We assume that the only ions contributing to FFA are H$^+$, He$^+$ and He$^{++}$. For a species $i$ with charge $Ze$ the linear absorption coefficient at a frequency $\nu$ is
\begin{equation}
\alpha_{\nu,i}(Z) = \sqrt{\frac{32 \upi}{27}} c^2 r_0^3 \left( \frac{kT}{m_ec^2} \right)^{-3/2} n_e n_i(Z) Z^2 g_\nu(T, Z) \nu^{-2}
\label{eq: linear absorption coefficient}
\end{equation}
where $n_i(Z)$ and $n_e$ are the species and electron number densities respectively, $T$ is temperature and $g(T, Z)$ is the Gaunt factor.
For clarity, we write
\begin{equation}
\alpha_{\nu,i}(Z) = n^2 K T^{-3/2} \chi_{\nu, i} (Z) \nu^{-2}
\end{equation}
where $K$ is a collection of constants and 
\begin{equation}
\chi_{\nu, i}(Z) = \frac{n_e}{n} \frac{n_i(Z)}{n} Z^2 g_\nu(T, Z).
\label{eq: chi}
\end{equation}
If the absorption coefficients $\alpha_\nu$ are constant along the line of sight, then the expected FFA optical depth is given by
\begin{equation}
\tau_\nu = E(n^2) K T^{-3/2} \left[ \sum_{i}\sum_{Z} \chi_{\nu, i} (Z) \right] \nu^{-2} L
\label{eq: FFA optical depth tau}
\end{equation}
where $E(n^2)$ is the expected value of $n^2$ given the density PDF, and $L$ is the depth of the absorbing screen.
We assume that the 100\,pc-scale jets ionize clouds of material on the inner edge of the circumnuclear disc, which then free-free absorb radio emission from the jet plasma. This is consistent with the high-velocity \hi{} clouds and regions of free-free absorbed 1.4\,GHz continuum emission detected in front of both lobes\,\citep{Conway1996}. Hence we take $L = 100\,\rm pc$.

The parameter $s$ in the log-normal distribution can be related to the properties of the ISM using
\begin{equation}
s^2 = \ln\left(1 + b^2 \mathcal{M}^2 \frac{\beta}{\beta + 1}\right)
\label{eq: s^2 in log normal distribution}
\end{equation}
which applies when the magnetic field strength $B \propto \rho^{1/2}$\,\citep{Federrath2012}. Here, $b$ is the turbulent forcing parameter, $\beta$ is the ratio of thermal to magnetic pressure and $\mathcal{M} = \sigma_{v} / c_s$ is the Mach number where $\sigma_v$ is the velocity dispersion and $c_s = \sqrt{kT / \mu^\prime \rm amu}$ is the sound speed.
We assume the velocity dispersion of the absorbing medium is that of the \feii{} emission, approximately $350\,\rm km\,s^{-1}$.

We calculate $\mu$ and $\sigma^2$ as follows, using the input parameters shown in Table~\ref{tab: parameters used in ISM calculation}.
First, we use a MAPPINGS\,V\,\citep{Sutherland2013} model grid with non-equilibrium cooling and solar abundances to compute the fractional abundances of electrons, H$^+$, He$^+$ and He$^{++}$ at a temperature $T \approx 10^4\,\rm K$. 
For each species, we use these values to calculate $\chi_{\nu, i}(Z)$ using Eqn.\,\ref{eq: chi}.
Then, by setting $\tau_\nu = 1$ at the spectral peak ($400\,\rm MHz$) in Eqn.\,\ref{eq: FFA optical depth tau}, we solve for $E(n^2)$. 
We then find $s$ using Eqn.\,\ref{eq: s^2 in log normal distribution}, which in turn allows us to find $m$ using using Eqn.\,\ref{eq: E(n2) of log-normal distribution}.  Finally, we solve for $\mu$ and $\sigma$ using Eqns.\,\ref{eq: E(n) of log-normal distribution} and \ref{eq: sigma of log-normal distribution}, yielding $\mu = 15.5\,\rm cm^{-3}$ and $\sigma^2 = 1.86 \times 10^{4}\,\rm cm^{-6}$. 

\begin{table}
	\caption{Parameters used in determining the parameters of the log-normal density distribution. Output parameters are denoted with daggers ($^\dagger$).}
	\begin{tabular}{c c c}
		\hline
		\textbf{Parameter} & \textbf{Symbol} & \textbf{Value} \\
		\hline
		Peak frequency & $\nu_{\rm peak}$ & $400\,\rm MHz$ \\
		Depth of absorbing slab & $L$ & $100\,\rm pc$\\
		Temperature & $T$ & $10059\,\rm K$ \\
		Mean molecular mass & $\mu^\prime$ & 0.66504 \\ 
		Electron fractional abundance & $n_e / n$ & $0.47175$ \\
		H$^+$ fractional abundance & $n_{\rm H^+} / n$ & $0.41932$ \\
		He$^+$ fractional abundance & $n_{\rm He^+} / n$ & $0.024458$ \\
		He$^{++}$ fractional abundance & $n_{\rm He^{++}} / n$ & $0.013770$ \\
		Velocity dispersion & $\sigma_v$ & $350\,\rm km\,s^{-1}$ \\
		Turbulent forcing parameter & $b$ & $0.4$ \\
		Ratio of thermal to magnetic pressure & $\beta$ & $1$ (equipartition) \\
		\hline
		Expected value of $n^2$$^\dagger$ & $E(n^2)$ & $1.89 \times 10^{4}\,\rm cm^{-6}$\\
		Mean density$^\dagger$ & $\mu$ & $15.5\,\rm cm^{-3}$ \\
	 	Density variance$^\dagger$ & $\sigma^2$ & $ 1.86 \times 10^{4}\,\rm cm^{-6}$ \\
		\hline
	\end{tabular}
	\label{tab: parameters used in ISM calculation}
\end{table}

%%%%%%%%%%%%%%%%%%%%%%%%%%%%%%%%%%%%%%%%%%%%%%%%%%
\section{Conclusion}\label{sec:Conclusion}

We have reported $H$- and $K$-band Gemini/NIFS observations of the Compact Steep Spectrum source 4C\,31.04, a young AGN with jets interacting strongly with a clumpy ISM. 
The host of 4C\,31.04 is a $z = 0.0602$ elliptical galaxy that harbours $\sim 10^{9}\rm \,\rm M_\odot$ of gas in a circumnuclear disc $\approx 2\rm\,kpc$ in diameter.

In the central few $100\rm\,pc$, we detect \feii{} emission that has a radial velocity field consistent with an expanding bubble driven by the jets.
We rule out SNe explosions as the cause of the \feii{} emission; moreover, the kinematics of the line trace an expanding bubble, implying the emission is a result of a jet-ISM interaction. 
 
We also detect ro-vibrational \hh{} emission that traces $\sim 10^4\,\rm M_\odot$ of very warm ($\sim 10^3\,\rm K$) \hh{}. This warm molecular phase traces rapidly cooling gas in the innermost $\sim \rm kpc$ of the circumnuclear disc, and represents a small fraction of a much larger ($\sim 10^6\,\rm M_\odot$) reservoir of warm ($\sim 10^2\,\rm K$) \hh{}. 
Near- and mid-IR line ratios indicate both \hh{} components are excited by shocks.
We show that shocks driven by accretion of gas on to the kpc-scale circumnuclear disc is unable to reproduce the observed \hh{} luminosity, and conclude that the shocks must be driven by jet plasma percolating to kpc radii through channels in the clumpy disc. 
The $\sim 10^3\,\rm K$ \hh{} emission shows a systemic blueshift of $\approx 150\,\rm km\,s^{-1}$ relative to the most widely accepted redshift of \citetalias{GarciaBurillo2007}. We speculate that the blueshift is caused by jet plasma radially accelerating clouds in the disc plane to kpc radii. Previous observations of 4C\,31.04 have revealed spatially-unresolved \hco{} and \hi{} in absorption at similarly blueshifted velocities, which may represent cooler gas entrained in the same low-speed outflow. The blueshift could also be explained if the redshift of the galaxy is in fact lower than currently believed, although this would impart a significant redshift to our \feii{} observations which would be difficult to explain. 

Our observations demonstrate that 4C\,31.04 is currently in the `flood-and-channel' phase of evolution that has been predicted by hydrodynamical simulations\,\citep[e.g.,][]{Mukherjee2016,SutherlandBicknell2007}, in which streams of jet plasma follow paths of least resistance through the ISM and form an energy-driven bubble. The bubble pushes a forward shock into the ambient ISM, giving rise to the \feii{} emission. Jet plasma also percolates into the circumnuclear disc, shocking and radially accelerating gas clouds, giving rise to the \hh{} emission.  

The extent of the shocked gas in our NIFS observations is much larger than the radio lobes resolved in VLBI imaging, suggesting the presence of low surface brightness radio plasma that has not been previously observed. This is consistent with multi-frequency VLBI observations of 4C\,31.04 with $< 100\rm\,per\,cent$ flux completeness. 
In simulations of jets propagating through clumpy discs, the brightest regions of plasma are temporarily halted by dense clumps, whilst the lower density plasma channels can continue to expand freely out of disc plane, enabling the bubble to grow much larger than the jets resolved by low-dynamic range VLBI observations. 

We estimated the jet flux using VLBI observations and use the observed bubble radius to constrain the `true' age of the radio jets. We find the jet age $\gtrapprox 17\,\rm kyr$, much older than previous literature estimates derived from the lobe expansion rate and using synchrotron spectral decay\,\citep{Giroletti2003}. 

Finally, we calculated the parameters of the density distribution of the ISM from the peak of the radio spectrum using a FFA model. These parameters together with our estimates of the jet flux will be used to inform future hydrodynamical systems tailored to 4C\,31.04.

Our observations of 4C\,31.04 demonstrate that optical and near-IR studies of radio galaxies can be crucial in estimating determining the true extent of the radio plasma, particularly in the early stages of evolution in which low surface brightness radio plasma may be resolved out by VLBI observations.

%%%%%%%%%%%%%%%%%%%%%%%%%%%%%%%%%%%%%%%%%%%%%%%%%%
\section*{Acknowledgements}

Based on observations obtained at the Gemini Observatory (processed using the \textsc{Gemini IRAF} package), which is operated by the Association of Universities for Research in Astronomy, Inc., under a cooperative agreement with the NSF on behalf of the Gemini partnership: the National Science Foundation (United States), the National Research Council (Canada), CONICYT (Chile), Ministerio de Ciencia, Tecnolog\'{i}a e Innovaci\'{o}n Productiva (Argentina), and Minist\'{e}rio da Ci\^{e}ncia, Tecnologia e Inova\c{c}\\,{a}o (Brazil). 

Based on observations made with the NASA/ESA \textit{Hubble Space Telescope}, and obtained from the Hubble Legacy Archive, which is a collaboration between the Space Telescope Science Institute (STScI/NASA), the Space Telescope European Coordinating Facility (ST-ECF/ESA) and the Canadian Astronomy Data Centre (CADC/NRC/CSA).

This research has made use of the NASA/IPAC Extragalactic Database (NED), which is operated by the Jet Propulsion Laboratory, California Institute of Technology, under contract with the National Aeronautics and Space Administration.

This work was supported by computational resources provided by the through the National Computational Infrastructure (NCI) facility under the National Computational and ANU Merit Allocation Schemes. 

The authors also wish to recognize and acknowledge the very significant cultural role and reverence that the summit of Mauna Kea has always had within the indigenous Hawai'ian community. 

The authors thank the reviewer for their insightful comments, and Matthew Colless and Alice Quillen for helpful discussions.

%%%%%%%%%%%%%%%%%%%%%%%%%%%%%%%%%%%%%%%%%%%%%%%%%%



\bibliographystyle{mnras}
\bibliography{bibliography} % if your bibtex file is called example.bib

%%%%%%%%%%%%%%%%%%%%%%%%%%%%%%%%%%%%%%%%%%%%%%%%%%

%%%%%%%%%%%%%%%%%% APPENDICES %%%%%%%%%%%%%%%%%%%%%

% Don't change these lines
\bsp	% typesetting comment
\label{lastpage}
\end{document}